\documentclass[10pt, a4paper]{article}
\usepackage{cite}
 \usepackage[utf8]{inputenc}
\usepackage{amssymb}
\usepackage{amsmath}
\usepackage{latexsym}
\usepackage{tikz}
\usepackage{sidecap,epstopdf}
\usepackage{subfigure}
\usepackage{color}
\usepackage{enumerate}
\usepackage{boondox-cal}
\usepackage{amsthm}\theoremstyle{remark}

\newcommand{\bte}{\begin{quote}\begin{theorem}}
\newcommand{\ete}[1]{\label{#1}\end{theorem}\end{quote}}
\newcommand{\bcom}{\begin{quote}\end{quote}}
\newcommand{\bex}{\begin{quote}\begin{example}}
\newcommand{\eex}[1]{\label{#1}\end{example}\end{quote}}
\newcommand{\bcon}{\begin{quote}\begin{conclusion}}
\newcommand{\econ}[1]{\label{#1}\end{conclusion}\end{quote}}
\newcommand{\bdefi}{\begin{quote}\begin{definition}}
\newcommand{\edefi}[1]{\label{#1}\end{definition}\end{quote}}

\newcommand{\blem}{\begin{quote}\begin{lemma}}
\newcommand{\elem}[1]{\label{#1}\end{lemma}\end{quote}}

\newcommand{\bpr}{\begin{quote}\begin{problem}}
\newcommand{\epr}[1]{\label{#1}\end{problem}\end{quote}}

\newcommand{\beq}{\begin{eqnarray}}
\newcommand{\eeq}[1]{\label{#1}\end{eqnarray}}

\newcommand{\bfi}{\begin{figure}[24]}
\newcommand{\efi}[1]{\caption{\label{#1}}\end{figure}}

\newcommand{\bfm}[1]{\mbox{\boldmath ${#1}$}}

\newcommand{\Ba}{{\textbf a}}
\newcommand{\Bb}{{\textbf b}}

\newcommand{\Be}{{\textbf e}}

\newcommand{\BA}{{\textbf A}}
\newcommand{\BB}{{\textbf B}}

\newcommand{\BF}{{\textbf F}}

\newcommand{\BI}{{\textbf I}}

\newcommand{\BL}{{\textbf L}}
\newcommand{\BM}{{\textbf M}}

\newcommand{\BQ}{{\textbf Q}}
\newcommand{\BR}{{\textbf R}}

\newcommand{\CA}{{\cal A}}

\newcommand{\CC}{{\cal C}}
\newcommand{\CD}{{\cal D}}

\newcommand{\CF}{{\cal F}}

\newcommand{\BGo}{\bfm\omega}

\newcommand\D{\,\mathrm{d}}
\newcommand\I{\mathrm{i}}
\newcommand\E{\mathrm{e}}
\newcommand{\bexe}{\begin{quote}\begin{exercise}\inh}
\newcommand{\eexe}[1]{\label{#1}\end{exercise}\end{quote}}

\usepackage{graphics,graphicx}
\usepackage{color}
\title{Vibrations and elastic waves in chiral multi-structures}

\author{M.J. Nieves$^{a,b,*}$, G. Carta$^c$, I.S. Jones$^c$, \\
A.B. Movchan$^d$ and N.V. Movchan$^d$ \\
\small{$^a$ School of Computing and Mathematics, Keele University, UK} \\
\small{$^b$ Department of Mechanical, Chemical and Materials Engineering,} \\
\small{University of Cagliari, Italy} \\
\small{$^c$Mechanical Engineering and Materials Research Centre,}\\
\small{ Liverpool John Moores University, UK} \\
\small{$^d$ Department of Mathematical Sciences, University of Liverpool, UK} \\
\small{$^*$Corresponding author; email address: m.nieves@keele.ac.uk} }

\date{}

\begin{document}
\maketitle
\begin{abstract}
\noindent We develop a new asymptotic model of the dynamic interaction between an elastic structure and  a system of gyroscopic spinners that make the overall multi-structure chiral. An important result is the derivation and analysis of effective chiral boundary conditions describing the interaction between an elastic beam and a gyroscopic spinner. These conditions are applied to the analysis of waves in systems of beams connected by gyroscopic spinners. A new asymptotic and physical interpretation of the notion of a Rayleigh gyrobeam is also presented. The theoretical findings are accompanied by illustrative numerical examples and simulations.
\end{abstract}
\section{Introduction}

Chirality,  the property of an  object whereby it is not congruent to its mirror image, occurs both through natural and man-made means in various areas of science.  The  useful and striking effects of chirality have received much attention in recent years, in particular in the development of optical metamaterials \cite{Wang2016}.
In mechanics, chirality may be introduced by gyroscopic spinners connected to a multi-structure, which may incorporate several elastic components. The present paper utilises  an asymptotic analysis to develop a new type of chiral boundary conditions and a subsequent study of a class of spectral problems for chiral elastic multi-structures.

The concept of chiral flexural elements, known as {\emph{gyrobeams}}, was  introduced in \cite{DEH1}. These chiral elements can be used for  controlling the attitude and shape of spacecraft during flight \cite{DEH2}. A gyrobeam can be interpreted as a beam with additional stored angular momentum whose effects are controlled by a spatial function governing the ``\emph{gyricity}"  of the element. This function allows for the coupling of the principal transverse motions in the beam. Several illustrations of the effect of gyricity  on the modes and stability of a beam have been presented in \cite{DEH1, DEH2, Yamanakaetal, HughesDEleuterio}.
Dynamic gyroelastic continuous models that utilise micropolar elasticity, have been developed in \cite{HassanHeppler1, HassanHeppler2} for one- and two-dimensional flexural media.

When waves propagate through civil engineering structures, such as bridges, pipeline systems and buildings, large deformations may occur  that can lead to the  collapse of the structure  \cite{Brunetal2}. Simplified discrete models offer ways in understanding possible vibration  \cite{Brunetal0, Giorgio5} and collapse \cite{Brunetal3, Nievesetal1, Nievesetal2} modes of such structures. These models may be easily adapted to include the effects of support systems capable of negating or re-routing the effects of unwanted vibrations generated, for instance, by seismic activity \cite{Giorgio4, Brunetal4}.
In addition, gyrobeams were recently used in a numerical model of seismic protection systems for civil engineering structures in \cite{Giorgio1a}, where a bridge support system composed of gyrobeams capable of diverting  low-frequency waves was proposed.

A new class of chiral boundary conditions has been introduced in \cite{Giorgio1} for a gyro-hinge connecting a gyroscopic spinner and an elastic beam.
Additionally in \cite{Giorgio1}, an infinite beam resting on a periodic distribution of gyro-hinges was used to approximate the low-frequency behaviour of a periodically supported gyrobeam with constant gyricity.
However, exactly how one quantifies the ``gyricity" of a gyrobeam  in terms of known mechanical quantities is an interesting question. This is addressed in the present paper.

Gyroscopic spinners have also found useful applications in the design of several two-dimensional chiral elastic structures that act as novel wave-guiding tools. Analysis of waves in a discrete triangular lattice whose nodes were attached to spinners was carried out in \cite{Brunetal1}.
In the time-harmonic regime, this lattice was shown to have novel filtering and polarising properties. In addition, the homogenised material associated with this lattice was used as an efficient cloaking and shielding device.  An in-depth analysis of the dispersive nature  and strong dynamic anisotropic properties of this chiral structure was carried out in \cite{Giorgio2}.
A heterogeneous arrangement of gyroscopic spinners can lead to surprising  wave propagation effects.
In particular, it has been shown in \cite{Giorgio3} that a triangular lattice, attached to two types of gyroscopic spinners,  admits waveforms localised in a single line, whose orientation can be controlled by adjusting the arrangement of spinners.

The approach developed in \cite{Brunetal1} was used in \cite{Wangetal,Garau} for numerical simulations of finite hexagonal systems resting on gyroscopic spinners  for the purpose of designing a robust topological insulator. 
Experimental evidence  demonstrating the topological insulation properties of a hexagonal lattice connected to gyroscopic spinners was given in \cite{Nashetal}.
Further examples where  chirality has been built into a discrete medium include systems of coupled pendula  that have been used to generate the mechanical analogue of the quantum Hall effect  \cite{SusstrunkHuber, Huber}, and tilted resonators that have been embedded in a triangular lattice to create localisation and interfacial waveforms \cite{Tallericoetal2,Tallaricoetal}.

Chirality can yield  counter-intuitive behaviour in the static response of a material. Examples of this include \cite{PrallLakes}, where honeycomb structures composed of rigid rings linked by slender ligaments were modelled and experimentally analysed.
Structures of this type are auxetic and their microstructure can be tuned to allow their effective behaviour to mimic a homogeneous material  possessing a negative Poisson's ratio. Wave propagation in such  structures has been investigated in \cite{Spadonietal}. A static micro-polar continuous model  has been used in \cite{Spadonietal2} as a homogenised representative of an auxetic material with a hexagonal microstructure. Homogenisation models for hexagonal and square cell chiral materials were presented in \cite{BacigalupoGambarotta}, where the dependency of the effective moduli on the underlying microstructural properties, including chirality, was examined. Chiral lattices containing inertial rings, capable of generating  low-frequency stop-bands for the structure, have been modelled in \cite{BacigalupoGambarotta2}.



In the present article, we  analyse chiral elastic multi-structures incorporating gyroscopic spinners connected to the end points of elastic beams. With the spinners being absent, the free ends of the beams would be subject to standard boundary conditions of zero shear forces and zero bending moments. However, with the spinners in place, a new set of chiral boundary conditions has been introduced to couple moments and hence rotational and flexural motion. Analysis of elastic waves in such systems incorporating chiral junctions is the main aim of the present study.


\begin{figure}
\begin{center}
\includegraphics[width=0.42\textwidth]{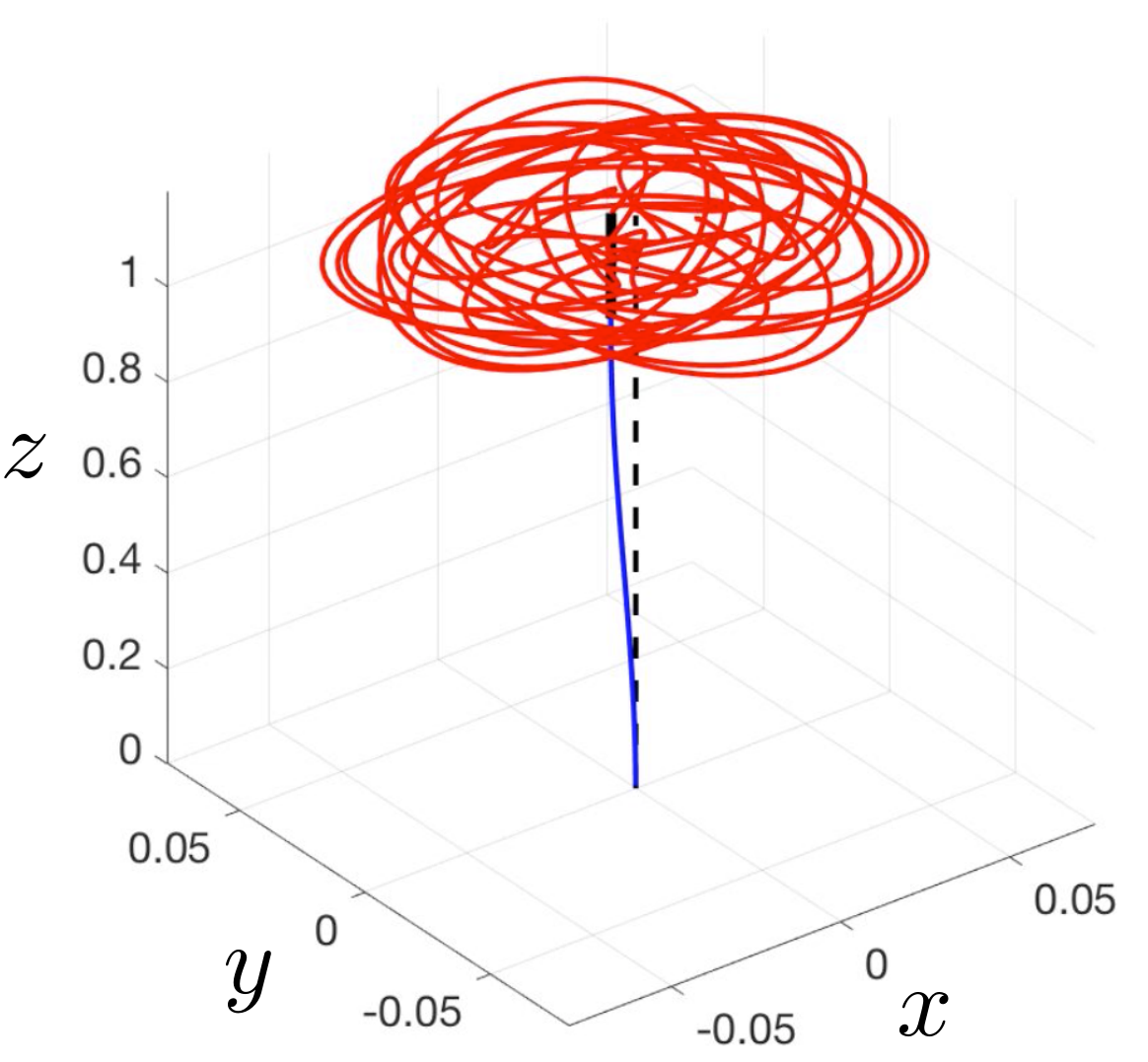} \qquad\qquad\includegraphics[width=0.4\textwidth]{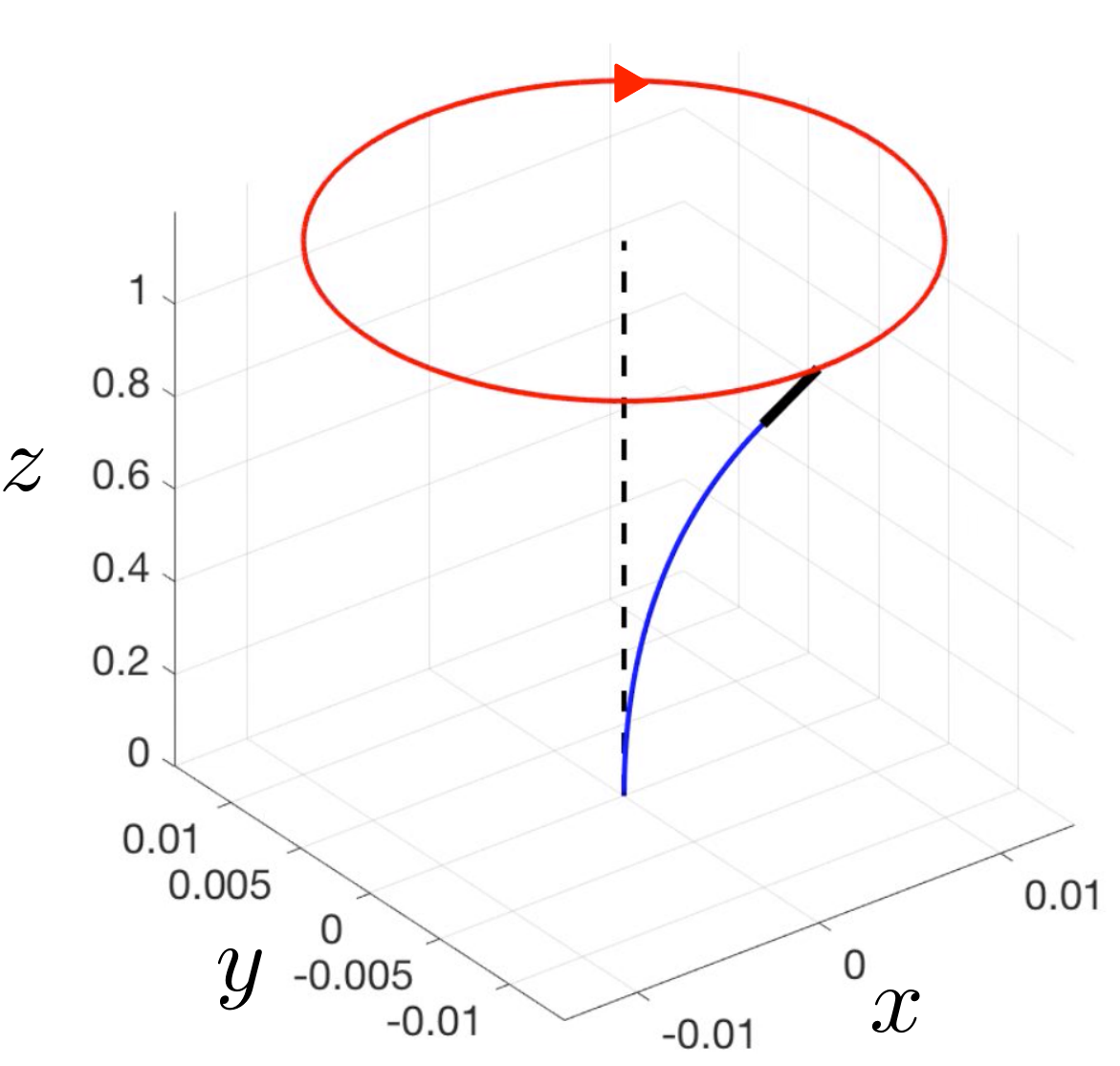}
(a)~~~~~~~~~~~~~~~~~~~~~~~~~~~~~~~~~~~~~~~~~~~~~~~~~~~~~(b)
\caption{Transient motion of an elastic massless beam connected to a gyroscopic spinner: (a) non-circular motion, (b) periodic circular motion. The motion of the system in (a) is a linear combination of four time-harmonic motions and an example of such a periodic motion is given in (b). The undeformed configuration is shown as the dashed line. The gyroscope is represented by the thick black line and the trajectory of the tip of the gyroscope is indicated by the contour at the top of the each diagram.}
\label{Fig0}
\end{center}
\end{figure}

Gyroscopic motion of a single gyroscope is described by a system of non-linear differential equations, well studied in the literature
(see, for instance, \cite{GPS}). When connected to a deformable solid, such as an elastic beam, a gyroscopic spinner produces a response which incorporates precession and nutation, combined with the elastic vibrations of the supporting structure.

In Figure \ref{Fig0} we show two examples of the transient behaviour of a multi-structure consisting of an elastic beam, clamped at the base, and a rotating gyroscopic spinner attached to the upper end of the beam (details are given in Section \ref{sec2EBgyro}).
Subject to the physical parameters of the system and to the initial conditions, the transient process may give different trajectories of the upper end of the beam. The example (a) in Figure \ref{Fig0} corresponds to a trajectory of a gyroscopic spinner placed on the top of a massless elastic beam. The pattern is a linear combination of four types of time-harmonic motions, as explained in Section \ref{Illustration}.
In some cases, an appropriate choice of initial conditions may lead to a periodic motion with a circular trajectory of the elastic multi-structure. An example of such a motion is shown in part (b) of  Figure \ref{Fig0}.
The direction of motion depends on the spin orientation of the gyroscopic spinner.
Such a periodic motion is well described by a class of eigenvalue problems for a chiral multi-structure, discussed formally in this paper.
In particular, a  linearised formulation will be discussed for the case when the angle of gyroscopic nutation is small.

The structure of the article is as follows.
In Section \ref{sec2EBgyro}, we present the formulation of the governing equations and  derive the chiral boundary conditions.
In Section \ref{THanalysis}, we analyse the time-harmonic motion of this system, and investigate how the presence of the gyroscopic spinner influences the eigenfrequencies and eigenmodes of the beam. In Section \ref{Periodicst}, we consider a system of beams connecting small, equally spaced gyroscopic spinners and we 
show that this system approximates a Rayleigh beam with an additional distribution of angular momentum.
Finally, in Section \ref{Conclusions}, we give some conclusions from the present work.








\section{Governing equations and chiral boundary conditions
}\label{sec2EBgyro}

\begin{figure}
\begin{center}
\includegraphics[width=0.25\textwidth]{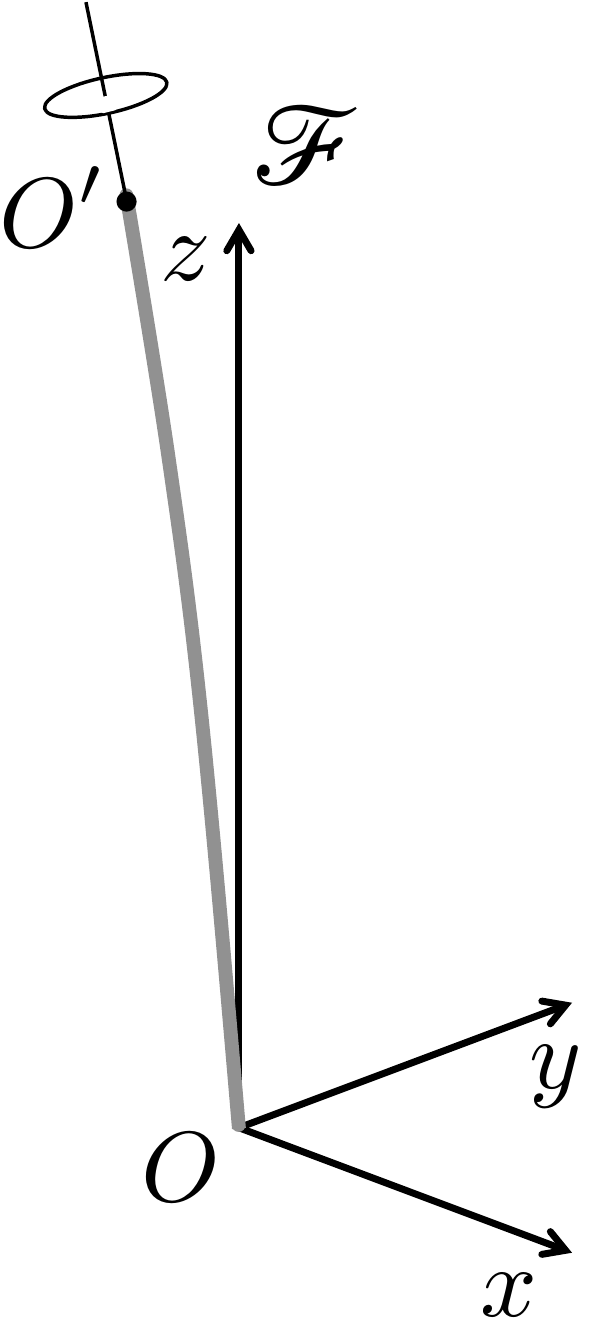} \qquad\qquad\includegraphics[width=0.44\textwidth]{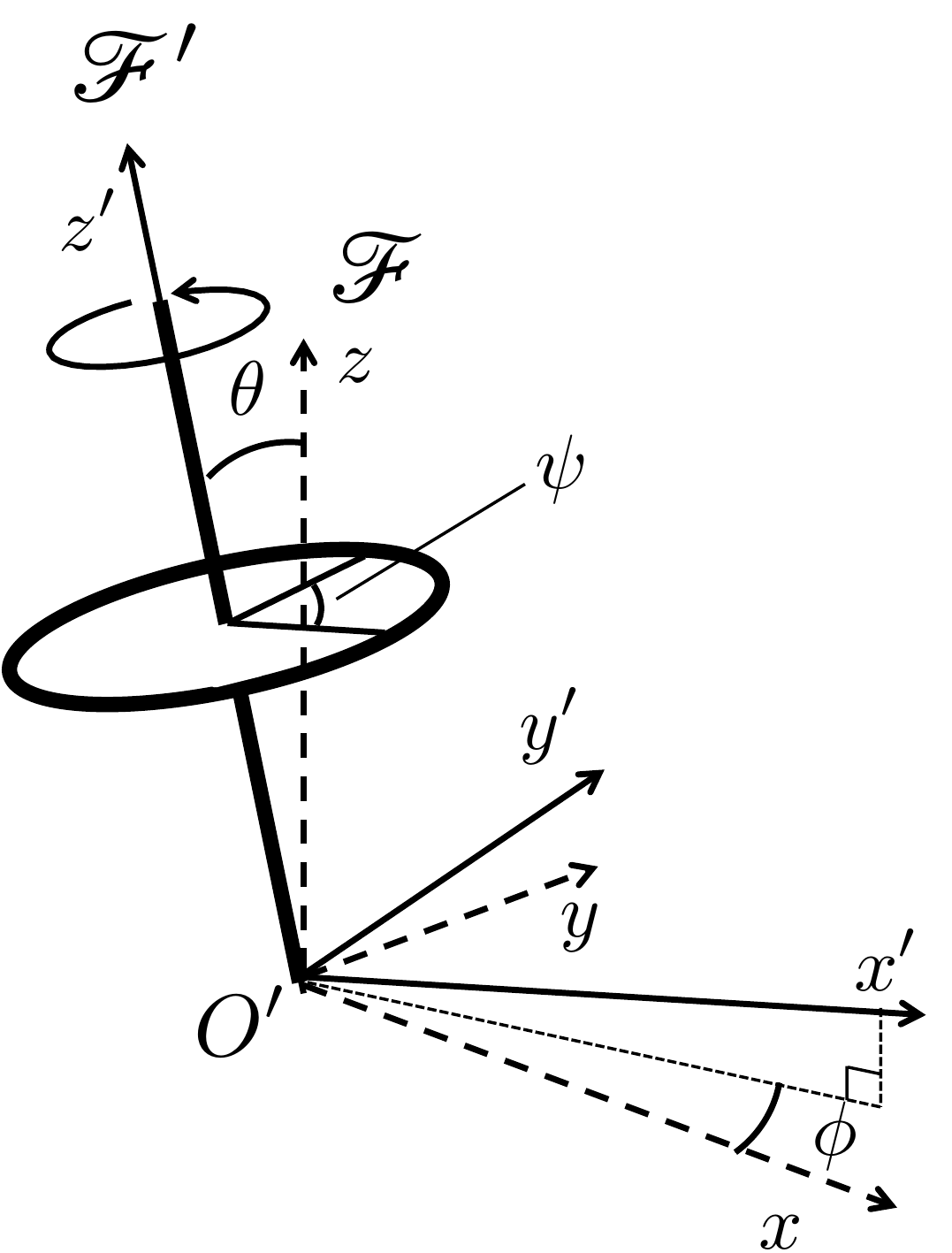}
(a)~~~~~~~~~~~~~~~~~~~~~~~~~~~~~~~~~~~~~~~~~~(b)
\caption{
(a) A beam (grey line) with a gyroscopic spinner connected to its tip. The beam is fixed at its base that is situated at the origin $O$ of a coordinate system $Ox y z$ in an inertial frame $\CF$.  (b) The gyroscopic spinner and the local coordinate system $O^\prime x^\prime y^\prime z^\prime$, which moves with the spinner as it precesses through an angle $\phi$ and nutates through an angle $\theta$. The coordinate system $O^\prime x^\prime y^\prime z^\prime$ is associated with the moving frame $\CF^\prime$ with origin $O^\prime$ at the base of the spinner.
}
\label{Fig1}
\end{center}
\end{figure}

We consider an Euler-Bernoulli beam, which is clamped at its base and is connected to a gyroscopic spinner at the other end (see Figure \ref{Fig1}(a)). It is assumed that the connection is such that the spinning motion of the spinner is not transmitted to the beam. We also assume that the fixture is such that the slope of the beam at the beam tip and the inclination of the spinner are the same at any time during the motion. In addition, we neglect the effect of gravity throughout.


\subsection{A chiral lower-dimensional model for an elastic beam
}\label{chiralbcs}
Let the beam have length $L$ and the beam's
 cross-section have identical
 second moments of area with respect to the $x$- and $y$-axes. In what follows the cross-section is assumed to be square.  Here $(x,y,z)$ denote the principal coordinates associated with an inertial frame $\CF$, whose origin coincides with the base of the beam.    The  coordinate in the direction of the beam's neutral axis is denoted by $z$, where $0\le z\le L$ (see Figure \ref{Fig1}(a)).
 Let $u(z,t)$, $v(z,t)$ and $w(z,t)$ be the displacement components in the $x$, $y$ and $z$ directions, respectively, of the beam at a point $z$
  and at time $t$.
 The displacements $u(z,t)$, $v(z,t)$ and $w(z,t)$ satisfy the following governing equations
 for $0< z< L$ and  $t>0$:
\begin{eqnarray}\label{eq1}
&& EJu^{\prime\prime\prime\prime}(z,t)+\rho A \ddot{u}(z,t)=0\;,\label{eq1b}\\
&& EJv^{\prime\prime\prime\prime}(z,t)+\rho A\ddot{v}(z,t)=0\;,\label{eq1c}\\
&& Ew^{\prime\prime}(z,t)-\rho\ddot{w}(z,t)=0\;,\label{eq1ca}
\end{eqnarray}
where $E$, $J$, $\rho$ and $A$ denote the Young's modulus, second moment of area, density and cross-sectional area of the beam, respectively. Here the prime and dot denote differentiation with respect to the spatial coordinate $z$ and
the time $t$, respectively.

At $z=0$, we assume that the beam is clamped:
\begin{eqnarray}\label{eq2a}
&&u(0, t)=v(0,t)=w(0,t)=0\;,\\
&& u^\prime(0, t)=v^\prime(0,t)=0\;.\label{eq2}
\end{eqnarray}

The formulation (\ref{eq1})--(\ref{eq2}) is accompanied by a set of boundary conditions at the top of the beam, which correspond to conditions  set at the connection between the tip of the beam and the base of the gyroscopic spinner. 
These conditions will be referred to as {\emph{chiral boundary conditions}}.

These effective boundary conditions represent the interaction between the elastic beam and the gyroscopic spinner. 
These boundary conditions reflect on the coupling between the displacements $u$ and $v$ and their derivatives and incorporate directional preference associated with the angular velocity of the gyroscopic spinner. It is also noted that the term ``chiral boundary conditions" can be extended to other types of elastic systems that involve a rotational preference. In the present paper, the term ``chiral boundary conditions" will be used consistently with reference to the effective junction conditions between an elastic beam and a gyroscopic spinner.

For a spinner having a small nutation angle, the boundary conditions representing the balance of forces (see (\ref{eq3}) and (\ref{eq3Fa})) are
\begin{eqnarray}\label{J1}
&&m\ddot{u}(L, t)+{mcl}
\ddot{u}^\prime(L,t)=EJu^{\prime\prime\prime}(L,t)\;,\\
&&m\ddot{v}(L, t)+{mcl}\ddot{v}^\prime(L,t)=EJv^{\prime\prime\prime}(L,t)\;,\label{J1a}\\
&&m\ddot{w}(L, t)=-EA w^{\prime}(L, t)\;,\label{J1aa}
\end{eqnarray}
and the chiral boundary conditions that correspond to the balance of moments at the tip of the beam  (see (\ref{MOMENTS}))
are given by 
\begin{eqnarray}\label{angmomcond}
&&-I_0\ddot{u}^\prime(L,t)-I_1\Omega \dot{v}^\prime(L,t)=EJ u^{\prime\prime}(L,t)\;,\label{J2a}\\
&&-I_0\ddot{v}^\prime(L,t)+ I_1\Omega \dot{u}^\prime(L,t)=EJv^{\prime\prime}(L,t)\;. \label{J2}
\end{eqnarray}
Here $m$ and $l$ are the mass and length of the gyroscopic spinner, respectively. We consider an axisymmetric gyroscopic spinner, whose centre of mass lies on the symmetry axis at a distance $cl$ from its base, where $0<c<1$.
The quantities $I_0$ and $I_1$ are the moments of inertia of the spinner about the principal transverse and vertical axes, respectively, of the local system $Ox^\prime y^\prime z^\prime$ associated with the spinner (see Figure \ref{Fig1}(b)). The parameter $\Omega$ is called the ``\emph{gyricity}" of the spinner, representing the combination of the spinner's initial spin and precession rates as in \cite{Giorgio1}. The conditions (\ref{angmomcond}) and (\ref{J2}) show that the gyricity provides the coupling between the functions $u$ and $v$ at the tip of the beam.
The derivation of (\ref{J1})--(\ref{J2}) is given in 
 Section \ref{Dofcbcs}.

The  initial conditions are
\begin{equation}
\begin{array}{c}
\displaystyle{u(z, 0)=u^{(0)}(z),\>\>\>  \frac{\partial u}{\partial t}(z, 0)=u^{(1)}(z)\;,} \quad
\displaystyle{v(z, 0)=v^{(0)}(z), \>\>\>  \frac{\partial v}{\partial t}(z, 0)=v^{(1)}(z)\;,}\\ \\
\displaystyle{w(z, 0)=w^{(0)}(z), \>\>\>  \frac{\partial w}{\partial t}(z, 0)=w^{(1)}(z)\;,}
\end{array}
\label{ics}
\end{equation}
where $u^{(j)}$, $v^{(j)}$, $w^{(j)}$, $j=0, 1,$ are given functions. Here, $u^{(j)}$ and $v^{(j)}$, $j=0, 1$, and their first order derivatives with respect to $z$  satisfy the homogeneous conditions at $z=0$, in order to be consistent with equations (\ref{eq2a})--(\ref{eq2}).  Additionally $w^{(j)}$, $j=0,1$, are zero for $z=0$.

Moreover, as a result of the connection of the gyroscopic spinner to the beam at $z=L$ (discussed in more detail in Section \ref{LB}),  the functions  $u^{(j)}$ and $v^{(j)}$, $j=0, 1,$ are subject to the conditions
\begin{equation}  \label{initrot}
\begin{array}{c}
\displaystyle{\frac{\partial u^{(0)}}{\partial z}(L)=\theta(0) \sin(\phi(0))\;,\quad  \frac{\partial v^{(0)}}{\partial z}(L)=-\theta(0) \cos(\phi(0))\;, } \\ \\
\displaystyle{\frac{\partial u^{(1)}}{\partial z}(L)=\frac{\partial}{\partial t}(\theta(t) \sin(\phi(t))\Bigg|_{t=0}\;,\quad  \frac{\partial v^{(1)}}{\partial z}(L)=-\frac{\partial}{\partial t}(\theta(t) \cos(\phi(t))\Bigg|_{t=0}\;,}
\end{array}
\end{equation}
where $\phi$ and $\theta$ are two of the Euler angles used to define the motion of the gyroscopic spinner (see Figure \ref{Fig1}(b)).

\subsection{Derivation of the chiral boundary conditions}\label{Dofcbcs}

Here,  the chiral boundary conditions (\ref{J1})--(\ref{J2}) are derived  from first principles.
The equations of motion of the gyroscopic spinner representing the balance of linear and angular momentum are combined with the dynamic response of  the elastic beam.  Linearisation is carried out, based on the assumption that the angle of nutation is small.
\subsubsection{Linear momentum balance for the gyroscopic spinner}\label{LB}
In this derivation we use the inertial frame $\CF$ and the non-inertial frame $\CF^\prime$ as shown in Figure \ref{Fig1}.
With respect to the basis associated with the inertial frame $\CF$, the motion of the spinner can be determined through its precession $\phi(t)$, nutation $\theta(t)$ and spin $\psi(t)$ (see Figure \ref{Fig1}(b)).
We introduce a non-inertial frame $\CF^\prime$ with coordinate system $O^\prime x^\prime y^\prime z^\prime$, as shown in Figure \ref{Fig1}(b), with $O'$ at the tip of the beam and the $z^\prime$-axis lying along the symmetry axis of the spinner.
The frame $\CF^\prime$   undergoes translations determined by the displacements $u(L, t)$, $v(L,t)$ and $w(L,t)$. This frame also rotates through the motion of the spinner as the latter nutates and precesses. Note that this frame $\CF^\prime$ does not spin with the spinner.

The frame $\CF^\prime$ may be obtained by a rotation of coordinates applied to the frame $\CF$. Indeed, the frame $\CF^\prime$ is obtained first by applying a rotation of coordinates through $\phi$ anticlockwise about the $z$-axis, followed by a rotation  of coordinates in this new system through $\theta$ anticlockwise about the transformed $x$-axis.
Thus, the centre of mass  of the gyroscopic spinner at a given time $t$ can be stated with respect to the frame $\CF$ as
\begin{eqnarray}\label{COM}
\BR_g&=&\Big(c\,{l}\sin(\theta)\sin(\phi)+u(L,t)\Big)\Be_1+\Big(v(L,t)-c\,{l}\sin(\theta)\cos(\phi)\Big)\Be_2\nonumber\\
&&+\Big(c\,{l}\cos(\theta)+w(L,t)+L\Big)\Be_3\;,
\end{eqnarray}
where $\{\Be_1,\Be_2,\Be_3\}$ is the Cartesian basis set for $(x, y, z)$.

Due to the axial symmetry of the gyroscopic spinner and the fact that the axes in the frame $\CF^\prime$ are a set of principal axes, the moment of inertia tensor of the spinner is diagonal. Let  $I_0$ be the moment of inertia of the spinner about the $x^\prime$- and $y^\prime$-axes  and let $I_1$ be the moment of inertia about the axis of the spinner, all in frame $\CF^\prime$.

In $\CF^\prime$, we use the time-dependent basis $\{\Be_1^\prime, \Be_2^\prime, \Be_3^\prime\}$.
This basis can be written in terms of the basis $\{\Be_1, \Be_2, \Be_3\}$ as follows:
\begin{eqnarray}
&&\Be^\prime_1=\cos(\phi(t))\Be_1+\sin(\phi(t))\Be_2\;,\label{eb1}\\
&&\Be^\prime_2=-\cos(\theta(t))\sin(\phi(t))\Be_1+\cos(\theta(t))\cos(\phi(t))\Be_2+\sin(\theta(t))\Be_3\;,\\
&&\Be^\prime_3=\sin(\theta(t))\sin(\phi(t))\Be_1-\sin(\theta(t))\cos(\phi(t))\Be_2+\cos(\theta(t))\Be_3\;.\label{eb3}
\end{eqnarray}
{We use the vector $\BQ$ to represent the shear forces in the beam
\begin{equation}\label{eq3}
\BQ(z,t)=-EJ  u^{\prime\prime\prime}(z,t) \Be_1-EJ  v^{\prime\prime\prime}(z,t)\Be_2\;.
\end{equation}
Together with the axial force along the beam in the  $z$-direction the total internal force $\BF$  is
\begin{equation}\label{eq3Fa}
\BF(z,t)=\BQ(z,t)+EA w^\prime (z,t)\Be_3\;.
\end{equation}}
The balance of linear momentum for the gyroscopic spinner, with respect to its centre of mass, takes the form
\begin{equation}\label{eqlinmombalance}
-\BF(L,t)=m\BA_g\;,
\end{equation}
where the left-hand side represents the  forces applied to the spinner by the tip of the beam (see (\ref{eq3Fa})),    and $\BA_g$ is the linear acceleration of the centre of mass of the spinner. The latter relation holds in the inertial frame $\CF$.

By use of (\ref{COM}) the linear acceleration is given by 
\begin{equation}\label{eqAg}
\BA_g=A_g^{(1)}\Be_1+A_g^{(2)}\Be_2+A_g^{(3)}\Be_3,
\end{equation}
where
\begin{eqnarray*}
A_g^{(1)}&=& \ddot{u}(L, t)+c\,l [\ddot{\theta}\cos(\theta)\sin(\phi)-\dot{\theta}^2\sin(\theta)\sin(\phi)\nonumber  \\
&&+2\dot{\theta}\dot{\phi}\cos(\theta)\cos(\phi)+\ddot{\phi}\sin(\theta)\cos(\phi)-\dot{\phi}^2\sin(\theta)\sin(\phi)]\;,\\\nonumber\\
A_g^{(2)}&=& \ddot{v}(L, t)+c\,l [-\ddot{\theta}\cos(\theta)\cos(\phi)+\dot{\theta}^2\sin(\theta)\cos(\phi)\nonumber  \\
&&+2\dot{\theta}\dot{\phi}\cos(\theta)\sin(\phi)+\ddot{\phi}\sin(\theta)\sin(\phi)+\dot{\phi}^2\sin(\theta)\cos(\phi)]\;,
\end{eqnarray*}
and
\begin{eqnarray*}
A_g^{(3)}&=& \ddot{w}(L, t)-c\,l [\ddot{\theta}\sin(\theta)+\dot{\theta}^2\cos(\theta)]\;.
\end{eqnarray*}
Substitution of (\ref{eq3Fa})  and (\ref{eqAg})  into (\ref{eqlinmombalance}) gives the linear equations of motion:
\begin{eqnarray}
&&m\ddot{u}(L, t)+mc\,l [\ddot{\theta}\cos(\theta)\sin(\phi)-\dot{\theta}^2\sin(\theta)\sin(\phi)\nonumber  \\
&&+2\dot{\theta}\dot{\phi}\cos(\theta)\cos(\phi)+\ddot{\phi}\sin(\theta)\cos(\phi)-\dot{\phi}^2\sin(\theta)\sin(\phi)]=EJ u^{\prime\prime\prime}(L, t)\;, \nonumber\\ \label{Mot1}\\
&&m\ddot{v}(L, t)+mc\,l [-\ddot{\theta}\cos(\theta)\cos(\phi)+\dot{\theta}^2\sin(\theta)\cos(\phi)\nonumber  \\
&&+2\dot{\theta}\dot{\phi}\cos(\theta)\sin(\phi)+\ddot{\phi}\sin(\theta)\sin(\phi)+\dot{\phi}^2\sin(\theta)\cos(\phi)] =EJ v^{\prime\prime\prime}(L, t)\;,\nonumber \\ \label{Mot2}
\end{eqnarray}
and
\begin{eqnarray}
 && m\ddot{w}(L, t)-mc\,l [\ddot{\theta}\sin(\theta)+\dot{\theta}^2\cos(\theta)]=-EA w^{\prime}(L, t)\;. \label{Mot3}
\end{eqnarray}

\subsubsection*{Linearisation with respect to the nutation angle}
We assume that the nutation angle and its derivatives satisfy the conditions
\begin{equation*}\label{constraints}
\Big|\frac{\displaystyle{\D^j\theta(t)}}{\displaystyle{\D t^j}}\Big|\ll 1 \;,  \quad 0\le j \le 2\;.
\end{equation*}
Under these conditions, the left-hand sides of (\ref{Mot1})--(\ref{Mot3}) can be linearised, to leading order, giving the following equations:
\begin{eqnarray}
&&m\ddot{u}(L, t)+{mcl}\Big[(\ddot{\theta}-\theta\dot{\phi}^2) \sin(\phi)+(2\dot{\phi}\dot{\theta}+\ddot{\phi}\theta)\cos(\phi)\Big]
=EJu^{\prime\prime\prime}(L,t)\;, \nonumber \\\label{lineq1}\\
&&m\ddot{v}(L, t)+{mcl}\Big[-(\ddot{\theta}-\theta \dot{\phi}^2)\cos(\phi)+(2\dot{\phi}\dot{\theta}+\ddot{\phi}\theta)\sin(\phi)\Big]=EJv^{\prime\prime\prime}(L,t)\;,\nonumber \\\label{lineq2}
\end{eqnarray}
and
\begin{equation}
 m\ddot{w}(L, t)=-EA w^{\prime}(L, t)\;. \label{lineq3}
\end{equation}
Note that (\ref{lineq3})
corresponds to the equation of longitudinal motion of a mass at the end of an elastic rod. This demonstrates with (\ref{eq1ca}) that, within the linearised model, 
the motion of the beam in the $z$ direction is not affected by the spinner.

Consider the boundary conditions arising from the balance of linear momentum for the gyroscopic spinner.
According to \cite{Giorgio1}, the rotations of the beam at $z=L$ may be linked to the parameters $\theta$ and $\phi$ via
\begin{equation}\label{uprimevprime}
u^\prime(L, t)=\theta \sin (\phi) \quad \text{ and }\quad v^\prime(L,t)=-\theta\cos(\phi)\;.
\end{equation}
This is a consequence of the slopes of the gyroscopic spinner and of the beam being equal at their connection. Thus we have
\begin{eqnarray}
&&\dot{u}^\prime(L,t)=\dot{\theta}\sin(\phi)+\theta \dot{\phi}\cos(\phi)\;,\\
&&\ddot{u}^\prime(L,t)=\ddot{\theta}\sin(\phi)+2 \dot{\theta}\dot{\phi}\cos(\phi)+\theta \ddot{\phi}\cos(\phi)-\theta \dot{\phi}^2\sin(\phi)\;,
\end{eqnarray}
and
\begin{eqnarray}
&&\dot{v}^\prime(L,t)=-\dot{\theta}\cos(\phi)+\theta \dot{\phi}\sin(\phi)\;,\\
&&\ddot{v}^\prime(L,t)=-\ddot{\theta}\cos(\phi)+2 \dot{\theta}\dot{\phi}\sin(\phi)+\theta \ddot{\phi}\sin(\phi)+\theta \dot{\phi}^2\cos(\phi)\;. \label{ddotv}
\end{eqnarray}
The relations (\ref{uprimevprime})--(\ref{ddotv}) can be used in (\ref{lineq1})--(\ref{lineq2}) to obtain the linearised boundary conditions written in terms of the beam displacements, 
as
\begin{eqnarray*}\label{eq1int}
&&m\ddot{u}(L, t)+{mcl}
\ddot{u}^\prime(L,t)=EJu^{\prime\prime\prime}(L,t)\;,\\
&&m\ddot{v}(L, t)+{mcl}\ddot{v}^\prime(L,t)=EJv^{\prime\prime\prime}(L,t)\;,\label{eq1aint}
\end{eqnarray*}
together with (\ref{lineq3}). This completes the derivation  of  (\ref{J1})--(\ref{J1aa}). It is noted that these conditions do not couple the displacements of the beam. This coupling appears in the boundary conditions connected with the  angular momentum balance for the spinner at the beam tip  and they are developed in the next section.

\subsubsection{Angular momentum balance for the gyroscopic spinner}\label{AB}
Here we derive the linearised boundary conditions which correspond to the balance of angular momentum of the gyroscopic spinner.
This balance is given by
\begin{equation}\label{ANGB}
-\BM(L, t)=\dot{\BL}(t)\;,
\end{equation}
where
\[\BL(t)=\BI \BGo\;,\]
$\BI$ is the moment of inertia tensor given by the diagonal $3\times 3$ matrix
\begin{equation*}\label{IF}
\BI=\text{diag}\{I_0,I_0, I_1\}\;, \quad I_0, I_1>0\;,
\end{equation*}
and $\BGo$ is the angular velocity measured with respect to the inertial frame $\CF$.
{In the  frame $\CF$, the vector $\BM$ representing the internal bending moments in the beam is given by
\begin{equation}\label{MOMENTS}
\BM(z,t)=-EJ v^{\prime\prime}(z,t) \Be_1+EJu^{\prime\prime}(z,t)\Be_2\;.
\end{equation}
Note that the moment component in the ${\bf e\rm}_3$-direction is equal to zero, since the beam is not spinning about the $z$-axis.}

As shown in  \cite{GPS}, to derive the conditions (\ref{J2a})--(\ref{J2})  it is appropriate to write the quantities appearing in (\ref{ANGB}) with respect to the basis in $\CF^\prime$, since in this frame the inertia tensor is diagonal. The angular velocity $\BGo$ of the gyroscopic spinner
can be written in the following form
\begin{equation*}\label{trueANG}
\BGo=\dot{\theta}\Be_1^\prime+\dot{\phi}\sin(\theta)\Be_2^\prime+(\dot{\phi}\cos(\theta)+\dot{\psi}){\Be}_3^\prime \;.
\end{equation*}
Combining this with (\ref{ANGB}), the Euler angles $\{\phi, \theta, \psi\}$ satisfy (see  \cite{Giorgio1, GPS})
\begin{eqnarray*}\label{angb1}
&&-M^\prime_1=I_0 \ddot{\theta}+\dot{\phi}\sin(\theta)[-I_0\dot{\phi}\cos(\theta)+I_1(\dot{\phi}\cos(\theta)+\dot{\psi})]
\;,\\
&&-M^\prime_2=I_0[\ddot{\phi}\sin(\theta)+2\dot{\theta}\dot{\phi}\cos(\theta)]-I_1\dot{\theta}(\dot{\phi}\cos(\theta)+\dot{\psi})\;,
\\
&&-M^\prime_3=I_1(\ddot{\psi}+\ddot{\phi}\cos(\theta)-\dot{\phi}\dot{\theta}\sin(\theta))\;.\label{angb3}
\end{eqnarray*}
The left-hand sides in the preceding equations are the moments imposed by the tip of the beam on the gyroscopic spinner.
The quantities $M^\prime_j$, $1\le j \le 3$, are the components of the vector $\BM$ in (\ref{MOMENTS}) in the basis $\Be^\prime_j$, $1\le j \le 3$ (see (\ref{eb1})--(\ref{eb3})), that is $\BM=\sum_{j=1}^3M^\prime_j \Be_j^\prime$, where
\begin{eqnarray}
&&M_1^\prime=EJ\left[ -v^{\prime\prime}(L, t)\cos(\phi)+u^{\prime\prime}(L, t)\sin(\phi)\right]\;, \nonumber  \\
&&M_2^\prime=EJ\cos(\theta)\left[v^{\prime\prime}(L, t)\sin(\phi)+u^{\prime\prime}(L, t)\cos(\phi)\right]\;,\nonumber
\end{eqnarray}
and
\begin{eqnarray*}
M_3^\prime=EJ\sin(\theta)\left[-v^{\prime\prime}(L, t)\sin(\phi)-u^{\prime\prime}(L, t)\cos(\phi)\right]\;.
\end{eqnarray*}

The initial conditions for the Euler angles are
\[\phi(0)=\phi^{(0)}, \>\>\>\dot{\phi}(0)=\phi^{(1)}\;, \quad \theta(0)=\theta^{(0)}, \>\>\> \dot{\theta}(0)=\theta^{(1)},\]
and
\[\psi(0)=\psi^{(0)}, \>\>\>\dot{\psi}(0)=\psi^{(1)}.\]
The  conditions for the precession and nutation define the initial rotations of the beam at its tip and their angular velocities  (see  (\ref{ics})--(\ref{initrot})).

By repeating similar steps from \cite{Giorgio1}, after changing the basis from $\{{\bf e}_1^\prime, {\bf e}_2^\prime, {\bf e}_3^\prime\}$ to $\{{\bf e}_1, {\bf e}_2, {\bf e}_3\}$ and applying a  linearisation with respect to the small nutation angle, we can also obtain the boundary conditions for the moments at the tip of the beam in the form
\begin{eqnarray}
&&-I_0\ddot{u}^\prime(L,t)-I_1\Omega \dot{v}^\prime(L,t)=EJ u^{\prime\prime}(L,t)\;,\label{eq2inta}\\
&&-I_0\ddot{v}^\prime(L,t)+I_1\Omega  \dot{u}^\prime(L,t)=EJv^{\prime\prime}(L,t)\;,\label{eq2int}
\end{eqnarray}
where the  gyricity is
\[\Omega=\dot{\psi}+\dot{\phi}=\text{Const}\;.\]
This completes the derivation of  (\ref{J2a})--(\ref{J2}).
The boundary conditions (\ref{eq2inta})--(\ref{eq2int}) show that the gyricity couples the functions $u$ and $v$.
These conditions are the same as those  derived in \cite{Giorgio1} for the case of a \emph{gyro-hinge}, where the base of the gyroscopic spinner does not translate.

\subsection{Linearised transient motion of a massless beam connected to a gyroscopic spinner}\label{Illustration}

In this section, we study the transient motion of a massless beam, clamped at its base and connected to a gyroscopic spinner at its tip.
Since longitudinal vibrations of the beam are decoupled from transverse vibrations, the latter can be considered separately.

 For a massless beam ($\rho = 0$), the governing equations are given by (see (\ref{eq1b})--(\ref{eq1c}))
\begin{equation}\label{eqmassless}
u^{\prime\prime\prime\prime}(z,t)=0\;, \quad v^{\prime\prime\prime\prime}(z,t)=0\;,
\end{equation}
whose solutions are cubic functions of $z$. Using the boundary conditions (\ref{eq2a})--(\ref{eq2}) at the clamped end for $u$ and $v$, the solutions of (\ref{eqmassless}) take the form
\begin{equation}\label{soleqmassless}
u(z,t)=U_{1}(t)z^3+U_{2}(t)z^2\;, \quad v(z,t)=V_{1}(t)z^3+V_{2}(t)z^2\;.
\end{equation}
The time-dependent coefficients $U_{1}(t)$, $U_{2}(t)$, $V_{1}(t)$ and $V_{2}(t)$ can be expressed in terms of the displacements and rotations of the beam at the connection with the gyroscopic spinner, denoted as $u^{c}(t)=u(L,t)$, $v^{c}(t)=v(L,t)$ and $\theta_{x}^{c}(t)=-v^\prime(L,t)$, $\theta_{y}^{c}(t)=u^\prime(L,t)$ respectively. Accordingly, the functions (\ref{soleqmassless}) are given by
\begin{equation}\label{soleqmassless2}
\begin{split}
&u(z,t)=-\left( \frac{2u^{c}(t)}{L^3}-\frac{\theta_{y}^{c}(t)}{L^2} \right)z^3+\left( \frac{3u^{c}(t)}{L^2}-\frac{\theta_{y}^{c}(t)}{L} \right)z^2\;,\\
&v(z,t)=-\left( \frac{2v^{c}(t)}{L^3}+\frac{\theta_{x}^{c}(t)}{L^2} \right)z^3+\left( \frac{3v^{c}(t)}{L^2}+\frac{\theta_{x}^{c}(t)}{L} \right)z^2\;.\\
\end{split}
\end{equation}
Substituting (\ref{soleqmassless2}) into the chiral boundary conditions (\ref{J1})--(\ref{J1a}) and (\ref{J2a})--(\ref{J2}), we obtain
\begin{equation}\label{systemtransient}
\begin{split}
m L^3 \ddot{u}^{c}(t) + m c l L^3 \ddot{\theta}_{y}^{c}(t) + 12 EJ u^{c}(t) - 6 EJ L\theta_{y}^{c}(t) = 0 \;,\\
m L^3 \ddot{v}^{c}(t) - m c l L^3 \ddot{\theta}_{x}^{c}(t) + 12 EJ v^{c}(t) + 6 EJ L \theta_{x}^{c}(t) = 0 \;,\\
-I_0 L^2 \ddot{\theta}_{y}^{c}(t) + I_1 \Omega L^2 \dot{\theta}_{x}^{c}(t) + 6 EJ u^{c}(t) - 4 EJ L \theta_{y}^{c}(t) = 0 \;,\\
I_0 L^2 \ddot{\theta}_{x}^{c}(t) + I_1 \Omega L^2 \dot{\theta}_{y}^{c}(t) + 6 EJ v^{c}(t) + 4 EJ L \theta_{x}^{c}(t) = 0 \;,\\
\end{split}
\end{equation}
which is a system of four second-order ordinary differential equations in the variables $u^{c}(t)$, $v^{c}(t)$, $\theta_{x}^{c}(t)$ and $\theta_{y}^{c}(t)$.

The general solutions of (\ref{systemtransient}) can be expressed as linear combinations of four time-harmonic solutions:
\begin{equation}\label{superposition}
\begin{split}
{\bf{\mathcal{U}}}(t) &= c_1 {\bf{\mathcal{u}}}_1 \mathrm{e}^{\mathrm{i}\omega_1 t} + c_2 {\bf{\mathcal{u}}}_2 \mathrm{e}^{\mathrm{i}\omega_2 t} + c_3 {\bf{\mathcal{u}}}_3 \mathrm{e}^{\mathrm{i}\omega_3 t} + c_4 {\bf{\mathcal{u}}}_4 \mathrm{e}^{\mathrm{i}\omega_4 t} \\
& + c_5 \overline{{\bf{\mathcal{u}}}_1} \mathrm{e}^{-\mathrm{i} \omega_1 t} + c_6 \overline{{\bf{\mathcal{u}}}_2 } \mathrm{e}^{-\mathrm{i} \omega_2 t} + c_7 \overline{{\bf{\mathcal{u}}}_3} \mathrm{e}^{-\mathrm{i} \omega_3 t} + c_8 \overline{{\bf{\mathcal{u}}}_4} \mathrm{e}^{-\mathrm{i} \omega_4 t} \;, \\
\end{split}
\end{equation}
where ${\bf{\mathcal{U}}}(t) = \left( u^{c}(t), v^{c}(t), \theta_{x}^{c}(t), \theta_{y}^{c}(t) \right)^{\mathrm{T}}$ and the bar denotes the complex conjugate. In (\ref{superposition}), the frequencies $\pm \omega_j$, $j=1,\dots ,4$, are the roots of the characteristic equation
\begin{equation}\label{characteristiceq}
\begin{split}
&[I_0 L^4 m \omega^4-2EJL(6I_0+L(3cl +2l)m)\omega^2 +12 EJ^2]^2\\
&-[I_1L\omega(L^3m\omega^2-12EJ)\Omega]^2=0
\end{split}
\end{equation}
and ${\bf{\mathcal{u}}}_j$, $j=1,\ldots,4$, represent the corresponding eigenvectors. The eight coefficients $c_j$, $j=1,\ldots,8$, can be determined from the eight initial conditions
\begin{equation}\label{initialconditions}
\begin{split}
& u^{c}(0)=u^{c}_0 \;, \quad v^{c}(0)=v^{c}_0 \;, \quad \theta_x^{c}(0)=\theta^{c}_{x \,0} \;, \quad \theta_y^{c}(0)=\theta^{c}_{y \,0} \;, \\
& \dot{u}^{c}(0)=\dot{u}^{c}_0 \;, \quad \dot{v}^{c}(0)=\dot{v}^{c}_0 \;, \quad \dot{\theta}_x^{c}(0)=\dot{\theta}^{c}_{x \,0} \;, \quad \dot{\theta}_y^{c}(0)=\dot{\theta}^{c}_{y \,0} \;, \\
\end{split}
\end{equation}
where $u^{c}_0$, $v^{c}_0$, $\theta^{c}_{x \,0}$, $\theta^{c}_{y \,0}$, $\dot{u}^{c}_0$, $\dot{v}^{c}_0$, $\dot{\theta}^{c}_{x \,0}$ and $\dot{\theta}^{c}_{y \,0}$ are given values.

In order to obtain a motion of the system corresponding to a single time-harmonic mode, we choose the initial conditions to
be consistent with a linear combination of the corresponding eigenvector and its complex conjugate. For example, to obtain a time-harmonic mode of frequency $\omega_1$ we can choose $c_1=\overline{c_5}$ and take all the other coefficients in (\ref{superposition})
to be zero. Then the initial conditions (\ref{initialconditions}) are chosen to be consistent with the eigenvector ${\bf{\mathcal{u}}}_1$.
In general the motion of the system will not be periodic in time unless $\omega_j/\omega_k$, $j, k=1, \dots, 4,$ are rational.

\begin{figure}
\includegraphics[width=1\textwidth]{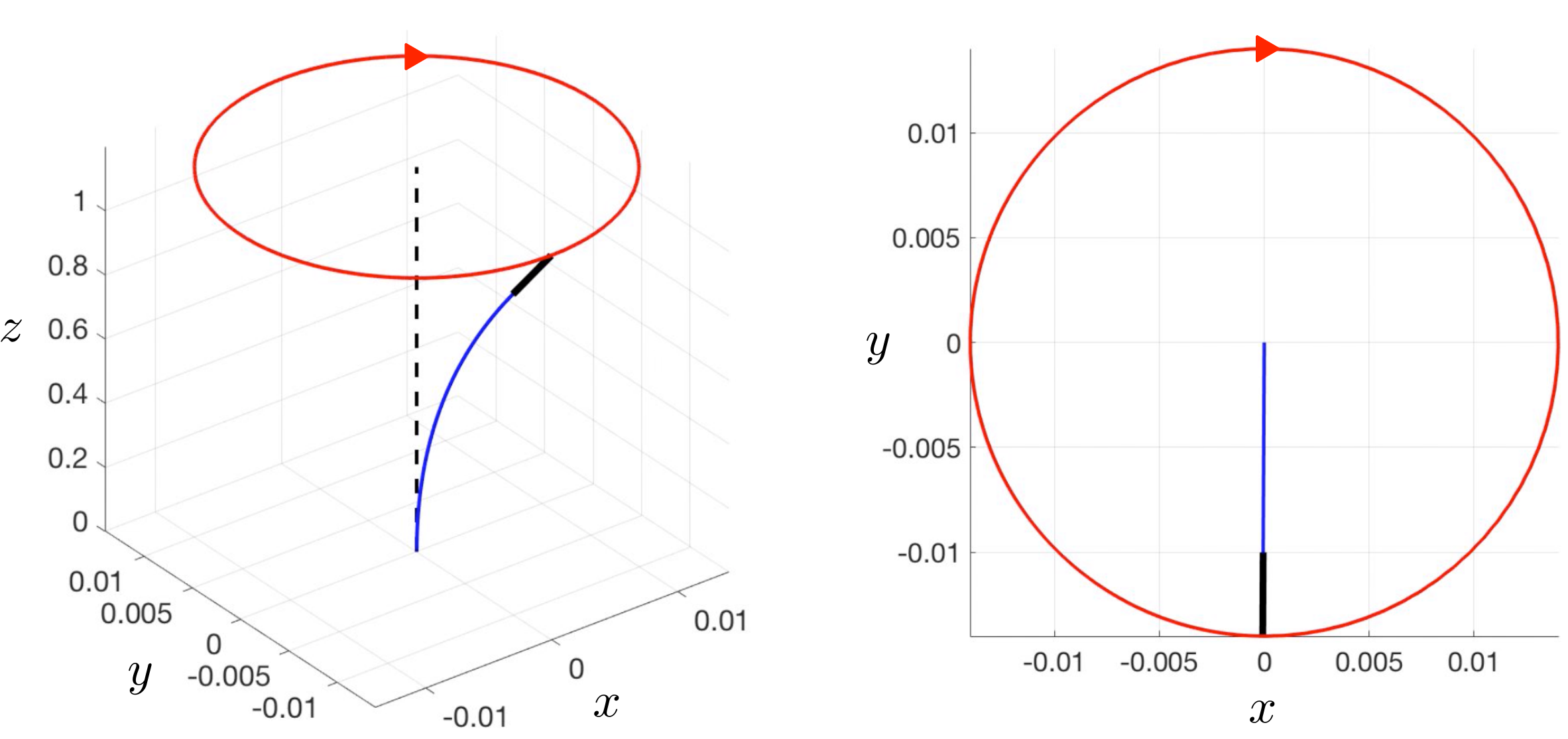}

~~~~~~~~~~~~~~~~~~~~~~~~~~(a)~~~~~~~~~~~~~~~~~~~~~~~~~~~~~~~~~~~~~~~~~~~~~~~~~~(b)
\caption{The behaviour of a massless beam, clamped at the base and with a gyroscopic spinner connected to its tip, as described in Section \ref{Illustration}. The computation is performed over a time interval of 100 s.
In (a) we show the profile of the beam  and the orientation of the spinner, indicated by the thick black line. The trajectory taken by  the tip of the spinner and the direction in which it traverses this path is shown by the circular contour and the arrow, respectively. In (b) we present the view of the structure from above.
}
\label{free_gyro_0}
\end{figure}

As an  illustrative example for a single time-harmonic mode of motion,
we take $L = 1$ m, $cl = 0.1$ m, $EJ = 1$ N$\,$m$^2$, $I_0 = 1$ kg$\,$m$^2$, $I_1 = 0.5$ kg$\,$m$^2$ and $\Omega = 15$ rad/s. In order to have $c_2 = c_3 = c_4 = c_6 = c_7 = c_8 = 0$, we impose the following initial values: $u^{c}_0 = 0$ m, $v^{c}_0 = -0.01$ m, $\theta^{c}_{x \,0} = 0.02$ rad, $\theta^{c}_{y \,0} = 0$ rad, $\dot{u}^{c}_0 = -0.00131$ m/s, $\dot{v}^{c}_0 = 0$ m/s, $\dot{\theta}^{c}_{x \,0} = 0$ rad/s and $\dot{\theta}^{c}_{y \,0} = -0.00262$ rad/s.
The results are shown in Figure \ref{free_gyro_0}. In panel (a),  it is possible to see the profile of the beam  and the trajectory of the top end of the spinner (shown by the circular contour at the top of the diagram).
A view of the system from above is presented in  panel (b).

From Figure \ref{free_gyro_0}, it can be seen that the tip of the beam moves in a
 circle with  radius $0.01$ m and centre at $(0, 0)$. The radius of the  trajectory coincides with the initial value of the displacement given to the beam tip.
The tip of the spinner follows a circular trajectory with centre at $(0,0)$ and radius approximately $0.015$ m.
The system takes $t=47.972$ s to complete a period of its motion, with this time corresponding to the radian  frequency $\omega=\omega_1=0.131$ rad/s.

The motion of the system can  also be observed in the Video 1 included in the Supplementary Material. The video shows that the entire system rotates clockwise and all points in the system move along a circular trajectory.

The above example confirms that the linearised transient motion of a gyroscopic spinner connected to an elastic massless beam is represented as a linear combination of four time-harmonic motions, and an appropriate choice of the initial conditions provides a periodic solution where the tip of the gyroscopic spinner moves along a circular trajectory.
The transient problem becomes more complicated when the beam has a non-zero mass density. In this case, a solution of the transient problem  is written as an infinite series where individual terms correspond to certain time-harmonic model problems.

In Section \ref{THanalysis}, we present the modal analysis of the chiral multi-structure for the more general case when the beam density is non-zero.

\section{Modal analysis for a beam connected to a gyroscopic spinner}\label{THanalysis}
Here, we concentrate on the transverse vibration modes for an inertial beam with a  fixed  base and with a gyroscopic spinner  connected to its tip in the time-harmonic regime.


\subsection{Normalised form of the boundary conditions}\label{normjunc}
We introduce normalisations to retrieve a dimensionless form of the boundary conditions.
This is carried out with the following change of  variables 
\begin{equation*}
 z=L \tilde{z}\;, \quad t=\sqrt{\frac{\rho A L^4}{EJ}} \tilde{t}\;,
 \end{equation*}
 and the normalisations
 \begin{equation*}u(z,t)=L  \tilde{u}(\tilde{z}, \tilde{t})\;, \quad v(z,t)={L}\tilde{v}(\tilde{z}, \tilde{t})\;,
\end{equation*}
\begin{eqnarray*}
&&I_1=\rho A L^3 \tilde{I_1}\;, \quad I_0=\rho A L^3 \tilde{I}_0\;.
\end{eqnarray*}
The eigenfrequencies of the multi-structure and the gyricity are then normalised by
\begin{equation*}
\omega=\tilde{\omega}\sqrt{\frac{EJ}{\rho A L^4}} \quad \text{and}  \quad \Omega=\sqrt{\frac{EJ}{\rho A L^4}} \tilde{\Omega}\;,
\end{equation*}
respectively.
We also introduce the parameters
\begin{equation*}
\alpha=\frac{m}{\rho A L}, \quad  \beta=\frac{l}{L}\;,
\end{equation*}
representing the mass  contrast between the beam and the spinner and the contrast between the length of the  spinner and the length of the beam, respectively. In addition, the parameter $\gamma$, given by
\begin{equation*}\label{gamma}
\gamma=\frac{\tilde{I}_1}{\tilde{I}_0}\;,
\end{equation*}
 represents the contrast in the principal moments of inertia for the spinner.
Using the above,  the normalised form of the boundary conditions (\ref{J1})--(\ref{J1a}) and (\ref{J2a})--(\ref{J2}) are expressed as
\begin{eqnarray}\label{eqn1}
&&{\alpha c \beta}
\ddot{u}^\prime(1,t)+
\alpha
\ddot{u}(1, t)=
u^{\prime\prime\prime}(1,t)\;,\\
&&{\alpha c \beta}\ddot{v}^\prime(1,t)+\alpha {\ddot{v}}{}(1, t)={v^{\prime\prime\prime}}{}(1,t)\;,\label{eqn2}
\end{eqnarray}
\begin{eqnarray}
&&I_0[-
\ddot{u}^\prime
(1,t)-\gamma\Omega
\dot{v}^\prime
(1,t)]=u^{\prime\prime}(1,t)\;,\label{eqn3}\\
&&I_0[\gamma\Omega
\dot{u}^\prime(1,t)-\ddot{v}^\prime(1,t)]=v^{\prime\prime}
(1,t)\;,\label{eqn4}
\end{eqnarray}
where the symbols ``tilde" have been omitted for ease of notation and will be omitted in the non-dimensional development in the sections \ref{timeharmonic}--\ref{EManalysis}.


\subsection{Transcendental equation for the eigenfrequencies}
\label{timeharmonic}
In this section, we consider the time-harmonic response of the system described in Section \ref{sec2EBgyro}.
In this case, we assume that  the complex displacements $u$ and $v$ take the form
\begin{equation*}
u(z,t)=U(z) e^{\I \omega t}\;, \quad v(z, t)=V(z) e^{\I \omega t}\;,
\end{equation*}
where (under the normalisation of Section \ref{normjunc}) the amplitudes  $U$ and $V$ are solutions of the equations
\begin{equation}\label{EBeq}
U^{\rm IV}(z)-\omega^2 U(z)=0\;, \quad V^{\rm IV}(z)-\omega^2 V(z)=0\;,  \quad 0\le z \le 1\;,
\end{equation}
which lead to their representations as
\begin{eqnarray}
&&U(z)=A_1 \cos(\sqrt{\omega} z)+A_2 \sin(\sqrt{\omega} z)+A_3 \cosh(\sqrt{\omega} z)+A_4 \sinh(\sqrt{\omega} z)\;,\nonumber \\
&&V(z)=B_1 \cos(\sqrt{\omega} z)+B_2 \sin(\sqrt{\omega} z)+B_3 \cosh(\sqrt{\omega} z)+B_4 \sinh(\sqrt{\omega} z)\;. \nonumber \\ \label{soln}
\end{eqnarray}
These amplitudes should also satisfy the clamped boundary conditions at $z=0$,
\begin{equation}\label{eqnm1}
\left(\begin{array}{c}U(0)\\
V(0)\end{array}\right)={\bf 0\rm}\;, \quad  \left(\begin{array}{c}
U^\prime(0)\\
V^\prime(0)\end{array}\right)={\bf 0\rm}\;,
\end{equation}
and the boundary conditions at $z=1$  (see (\ref{eqn1})--(\ref{eqn4}))
\begin{eqnarray}
&&\left(\begin{array}{c}
U^{\prime\prime\prime} (1)\\
V^{\prime\prime\prime} (1)
\end{array}\right)=-{\alpha c \beta \omega^2} \left(\begin{array}{c}U^\prime(1)\\ V^\prime(1)\end{array}\right)-\alpha \omega^2\left(\begin{array}{c}U(1)\\ V(1)\end{array}\right)\;,\label{eqnm2}
\end{eqnarray}
\begin{eqnarray}
&&\left(\begin{array}{c}U^{\prime\prime}(1)\\ V^{\prime\prime}(1)\end{array}\right)= I_0 \omega \left(\begin{array}{cc}\omega & -\I \gamma\Omega\\
 \I \gamma\Omega & \omega
\end{array}\right)\left(\begin{array}{c}U^\prime(1)\\ V^\prime(1)\end{array}\right)\;.
\label{eqn4n}
\end{eqnarray}

After satisfying the conditions (\ref{eqnm1}),
we use (\ref{eqnm2})--(\ref{eqn4n})  to derive a compact representation of equations for the 
remaining constants $A_1$, $A_2$, $B_1$ and $B_2$
\begin{equation}\label{eqDR1}
\left(\begin{array}{cc}\BA & \I \BB \\
-\I \BB & \BA \end{array}\right)\left(\begin{array}{c}\Ba\\
\Bb\end{array}\right)=\mathbf{0}\;,
\end{equation}
where
\begin{equation}
\Ba=(A_1, A_2)^{\rm T}\;, \quad \Bb=(B_1, B_2)^{\rm T}\;,
\end{equation}
and
\begin{equation*}
\BA=\omega\left(\begin{array}{cc}
-I_0 \omega^{3/2} T_2-T_3 & I_0 \omega^{3/2} T_3 -T_4\\ \\
-\omega^{1/2} T_4+\alpha \omega T_1 +{\alpha c\beta}\omega^{3/2} T_2&
\omega^{1/2} T_1+\alpha \omega T_2 -{\alpha c \beta}\omega^{3/2}T_3\end{array}\right)\;,
\end{equation*}
\begin{equation}
\BB=I_0\gamma\Omega \omega^{3/2}\left(\begin{array}{cc}
-T_2 &T_3\\
0 &0\end{array}\right)\;.\label{BBM}
\end{equation}
In the above, $T_j$, $1\le j \le 4$, are functions of $\omega$ only and
\begin{eqnarray*}\label{eqDR2}
&& T_1=\sin(\sqrt{\omega})-\sinh(\sqrt{\omega})\;, \quad T_2=\cos(\sqrt{\omega})-\cosh(\sqrt{\omega})\;,\\
&& T_3=\sin(\sqrt{\omega})+\sinh(\sqrt{\omega})\;, \quad T_4= \cos(\sqrt{\omega})+\cosh(\sqrt{\omega})\;.
\end{eqnarray*}
The transcendental equation linked to system (\ref{eqDR1})--(\ref{BBM}) can be derived in the form
\begin{equation}\label{DR}
[\CA(\omega)]^2
-[I_0\omega^{1/2}\gamma\Omega X(\omega)]^2=0\;,
\end{equation}
with
\begin{eqnarray}
\CA(\omega)&=&\text{det}(\BA)/(2\omega^{5/2}) \nonumber \\
&=&\alpha c\beta \omega\sin(\sqrt{\omega}) \sinh(\sqrt{\omega}) +I_0 \omega^{3/2} X(\omega)- Z(\omega)\;,\label{XY1a}\\ \nonumber \\
X(\omega)&=&\alpha \sqrt{\omega} [\cos(\sqrt{\omega})\cosh(\sqrt{\omega})-1]\nonumber \\
 &&+\cos(\sqrt{\omega})\sinh(\sqrt{\omega})+\cosh(\sqrt{\omega})\sin(\sqrt{\omega})
 \label{XY1}
\;,
\end{eqnarray}
and
\begin{eqnarray}
Z(\omega)
&=&1+\cos(\sqrt{\omega})\cosh(\sqrt{\omega})\nonumber \\
 &&+\alpha \sqrt{\omega} [\cos(\sqrt{\omega})\sinh(\sqrt{\omega})-\cosh(\sqrt{\omega})\sin(\sqrt{\omega})]\label{Z}\;.
\end{eqnarray}

\subsection{Variation of eigenfrequencies with gyricity}
\label{cylindergyro}

\begin{figure}
\centering
\includegraphics[width=0.7\textwidth]{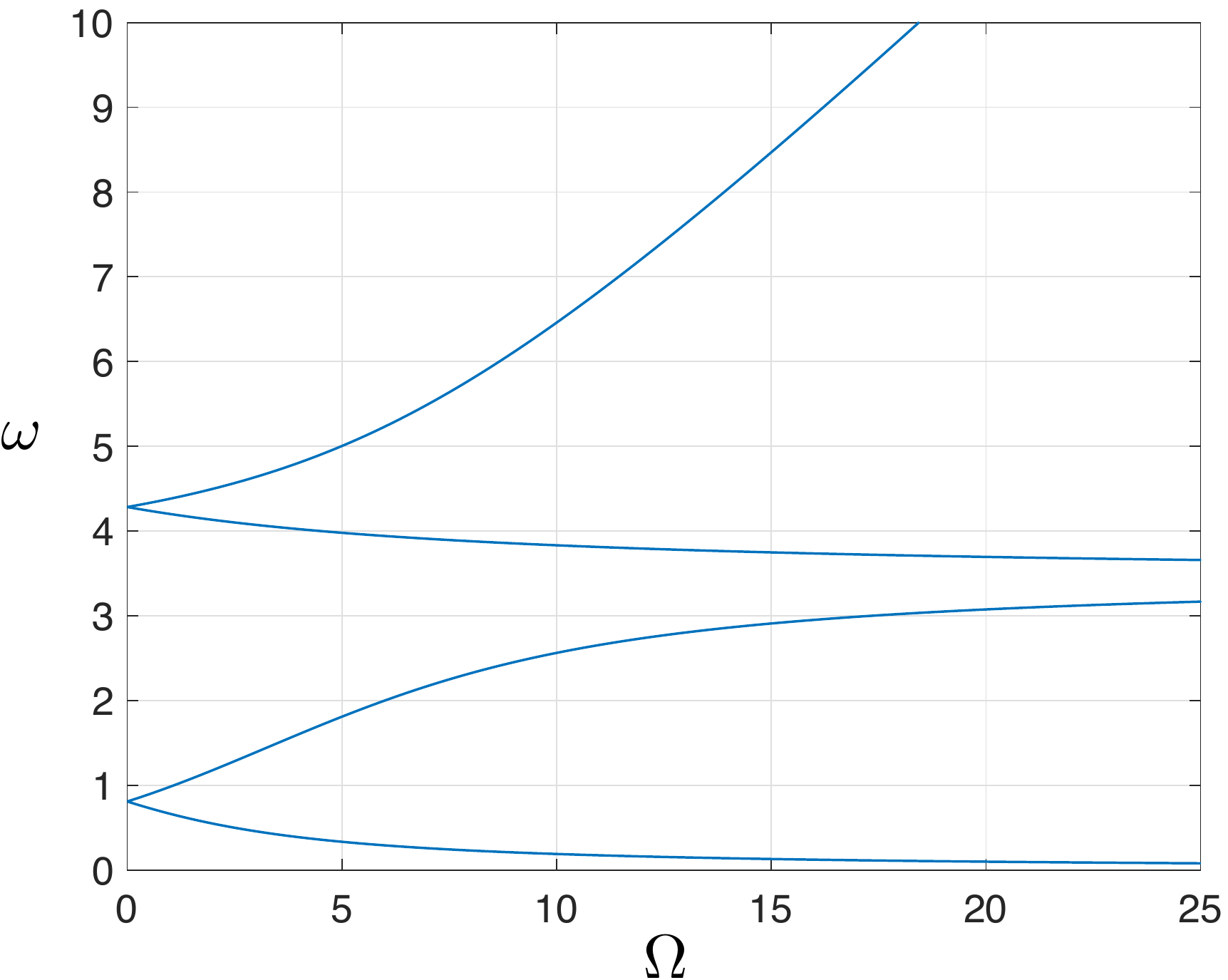}

\caption{Eigenfrequencies as functions of gyricity for a massless beam clamped at the base and with a gyroscope attached to its tip. Computations are performed for the parameters $c=1/2$, $I_1=0.5$$\text{ kg m}^2$,  $I_0=1\text{ kg m}^2$, 
$m=1\text{ kg}$, 
$L=1\text{ m}$,
$l=1\text{ m}$, $EJ=1\text{ Nm}^2$  and are based on (\ref{characteristiceq}). The corresponding normalised quantities introduced in Section \ref{normjunc} are $c=\gamma=1/2$, $I_0=\alpha=\beta=1$.
}
\label{fig_extra}
\end{figure}

\begin{figure}
\includegraphics[width=1\textwidth]{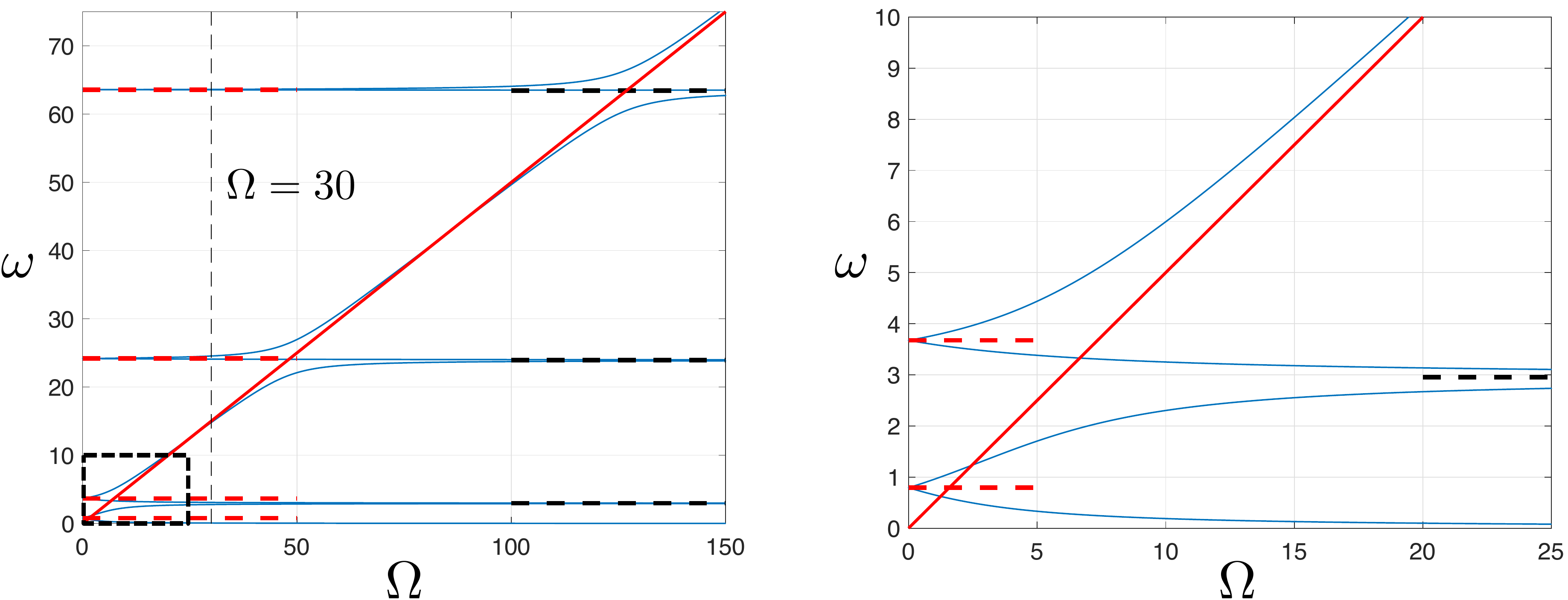}

~~~~~~~~~~~~~~~~~~~~~~~(a)~~~~~~~~~~~~~~~~~~~~~~~~~~~~~~~~~~~~~~~~~~~~~~~~(b)
\caption{(a) Eigenfrequencies of an inertial beam
with a clamped base at one end  and a gyroscopic spinner at the other end, as functions of gyricity. 
The parameter values are: $c=\gamma=1/2$, $I_0=\alpha=\beta=1$. 
The  line defined by $\omega=\gamma \Omega$ is also shown, which corresponds to the case of gyro-resonance discussed  in Section \ref{problem4_5}.
 (b) A magnification of the dashed box in part (a).
}
\label{free_gyro2}
\end{figure}

Figure \ref{fig_extra} shows the eigenfrequencies of a massless beam as functions of the gyricity, determined from (\ref{characteristiceq}).
For the sake of comparison, Figure \ref{free_gyro2} shows the eigenfrequencies of an inertial beam with the same properties as functions of the gyricity $\Omega$, obtained from (\ref{DR}). The values of the parameters are indicated in the captions of the figure.

In  Figures \ref{fig_extra} and \ref{free_gyro2}(a), we see that as the gyricity is increased, two branches emerge from a double eigenfrequency (indicated by the dashed line emanating from the left of Figure \ref{free_gyro2}(a)), which characterises the beam without spinner (see Section \ref{problem1}). One branch is a monotonic increasing function of the gyricity, while the other branch is a monotonic decreasing function of gyricity.
Both branches are bounded by the eigenfrequencies (indicated by the dashed lines on the right of Figure \ref{free_gyro2}) and these eigenfrequencies correspond to those of the problem discussed in Section \ref{problem2}.
The lower branch appears to be very flat in Figure \ref{free_gyro2}(a), but one can see the variation in the eigenfrequency as a function of $\Omega$ in Figure \ref{free_gyro2}(b), which provides a magnification of the branches within the dashed box of  Figure \ref{free_gyro2}(a).

It is noted that the first two curves in Figure \ref{fig_extra} are indistinguishable from those shown in Figure \ref{free_gyro2}(b). Indeed, according to the general theory of multi-scale elastic structures \cite{KMM}, within a finite range of frequencies adjacent to the origin, the eigenfrequencies of an inertial beam can be asymptotically approximated by those computed for the case of a massless beam.


\begin{figure}
\includegraphics[width=1\textwidth]{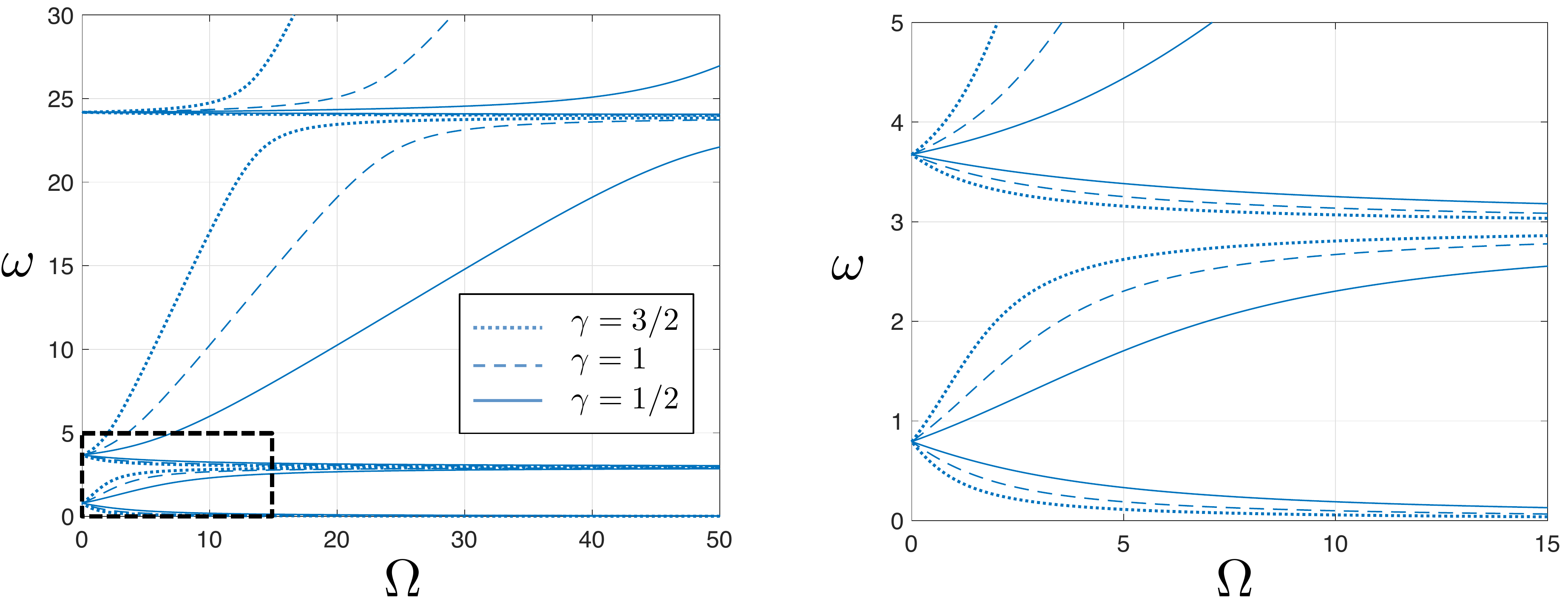}

~~~~~~~~~~~~~~~~~~~~~~~(a)~~~~~~~~~~~~~~~~~~~~~~~~~~~~~~~~~~~~~~~~~~~~~~~~(b)
\caption{(a) Eigenfrequencies as  functions of gyricity for an inertial beam with a clamped base and a gyroscopic spinner connected to its tip. The influence of $\gamma$ (the ratio of the principal moments of inertia of the spinner) on the behaviour of the eigenfrequencies is shown. The parameters are: $c=1/2$, $I_0=\alpha=\beta=1$, and $\gamma=1/2$, 1 and 3/2, represented  by the solid, dashed and dotted curves, respectively.  (b)  A magnification of the dashed box in Figure \ref{free_gyro4}(a).
}
\label{free_gyro4}
\end{figure}

\begin{figure}
\includegraphics[width=1\textwidth]{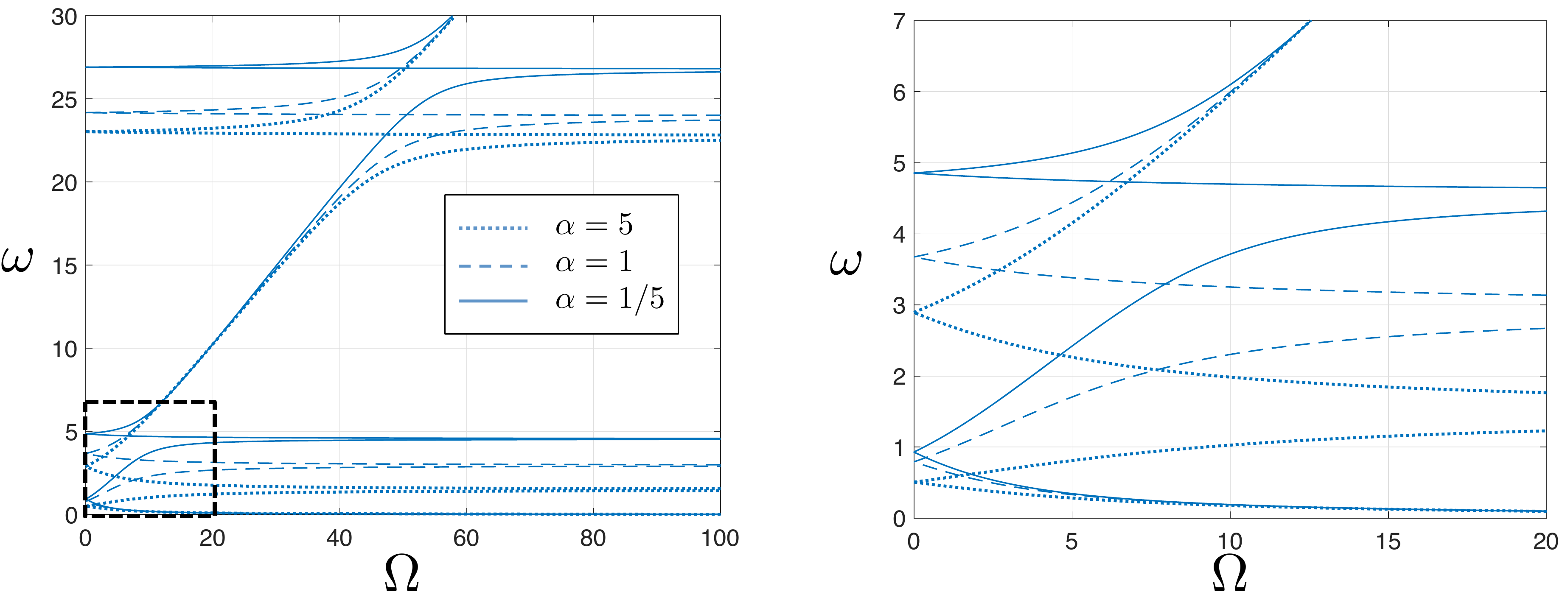}

~~~~~~~~~~~~~~~~~~~~~~~(a)~~~~~~~~~~~~~~~~~~~~~~~~~~~~~~~~~~~~~~~~~~~~~~~~(b)
\caption{(a) Eigenfrequencies as functions of gyricity for an inertial  beam with a clamped base and a gyroscopic spinner connected to its tip. The influence of  $\alpha$ (the mass contrast ratio between the gyroscopic spinner and the beam) on  the eigenfrequencies is shown. The parameters are: $c=\gamma=1/2$, $I_0=\beta=1$, and $\alpha=1/5$, 1  and 5, represented  by the solid, dashed and dotted curves, respectively. (b) A magnification of the   dashed box in Figure \ref{free_gyro3}(a).}
\label{free_gyro3}
\end{figure}

\begin{figure}
\includegraphics[width=1\textwidth]{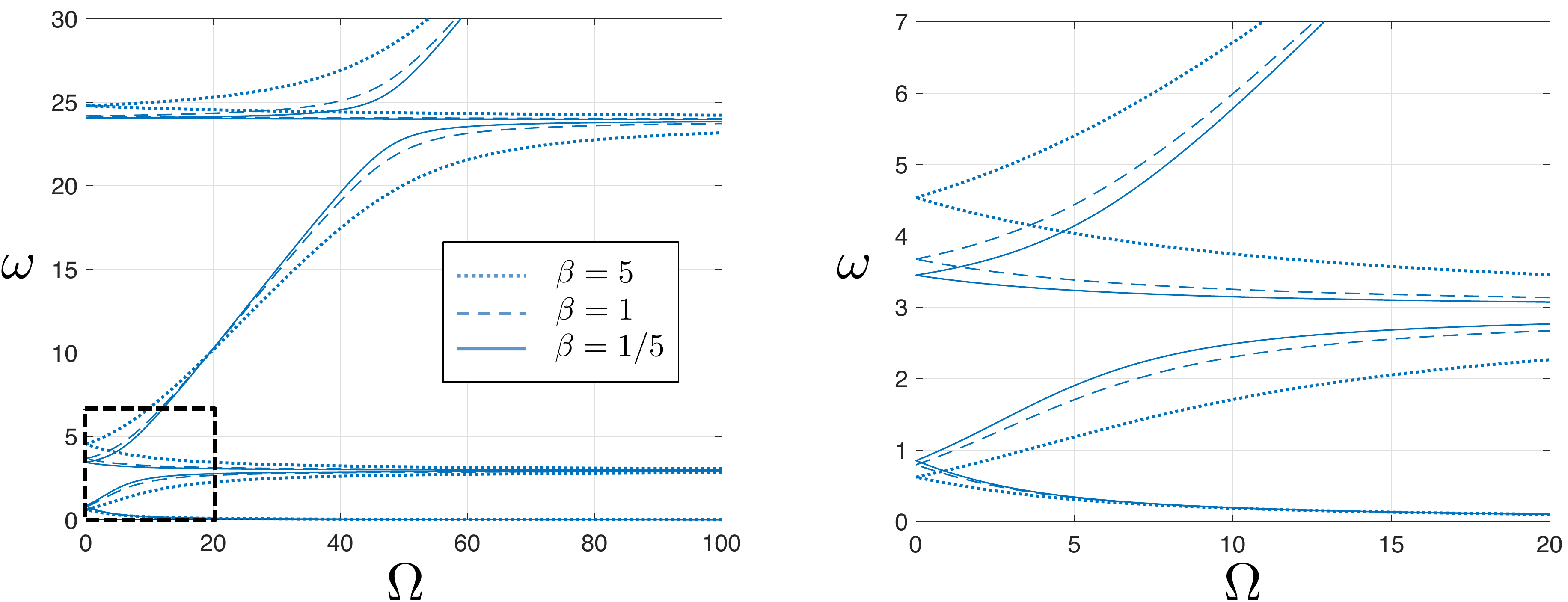}

~~~~~~~~~~~~~~~~~~~~~~~(a)~~~~~~~~~~~~~~~~~~~~~~~~~~~~~~~~~~~~~~~~~~~~~~~~(b)
\caption{(a) Eigenfrequencies as  functions of gyricity for an inertial beam with a clamped base and a gyroscopic spinner connected to its tip. The influence of  $\beta$ (the length contrast ratio between the spinner and the beam) on the eigenfrequencies is demonstrated.  The parameters are: $c=\gamma=1/2$, $I_0=\alpha=1$, and $\beta=1/5$, 1  and 5, represented  by the solid, dashed and dotted curves, respectively. (b) A magnification of  dashed box in Figure \ref{free_gyro5}(a).}
\label{free_gyro5}
\end{figure}

It is also of interest to investigate the influence of the parameters $\gamma$ (the contrast in the moments of inertia of the spinner),  $\alpha$ (the ratio of the mass of the spinner to the  beam's mass) and $\beta$ (the contrast in the length of the spinner to the beam's length) on the eigenfrequencies of the system.
Figures \ref{free_gyro4},  \ref{free_gyro3} and \ref{free_gyro5} show how changing the parameters $\gamma$,   $\alpha$ and $\beta$,  respectively,
affects the behaviour of the eigenfrequencies as functions of the gyricity. We note that, for all cases, the trends in the branches observed for $\Omega>0$  remain the same. We also mention that:

\begin{enumerate}[$\bullet$]
\item   the limits at $\Omega=0$, corresponding to a beam without the spinner (see Section \ref{problem1}), and the limit eigenfrequencies obtained when $\Omega \to \infty$ (see the problem of Section \ref{problem2}) are independent of $\gamma$. The variation of the eigenfrequencies for the system with gyricity, for different values of $\gamma$, can be seen in Figure \ref{free_gyro4}. As $\gamma$ increases the rate with which the upper and lower branches approach the limits for $\Omega \to \infty$ is also increased.

\item Figure \ref{free_gyro3} shows that increasing $\alpha$ lowers the values of the  limits obtained for $\Omega=0$ and $\Omega\to \infty$, as expected from  Sections \ref{problem1} and \ref{problem2}.   Increasing $\alpha$ corresponds to an increase in the mass of the gyroscopic spinner,  resulting in a decrease of the eigenfrequencies of the whole system. 

\item Figure \ref{free_gyro5} demonstrates that if $\beta$ is increased only the eigenfrequencies for the limits when $\Omega= 0$ increase (see the problem for zero gyricity in Section \ref{problem1}). On the other hand, the limit values for $\Omega \to \infty$ are independent of $\beta$, as shown in the problem  in this limit in Section \ref{problem2}.
 The rate with which the upper and lower branches converge to these values as $\Omega \to \infty$ decreases  with increase in $\beta$.
Similar effects can be observed if only the parameter $c$ governing the centre of mass is increased.

\end{enumerate}

\subsection{Special model problems}\label{spprob}


Here we outline several cases, where the boundary conditions at $z=1$  simplify or degenerate for a special choice of physical parameters and the values of the gyricity. In particular, we consider extremal cases when the gyricity is zero or large as well as cases of negligibly small moments of inertia. In the last part of the current section we address the state of a gyro-resonance, which is illustrated in the computational example in Figure \ref{free_gyro2}. The limit values of $\omega$ as a function of gyricity are used in Figures \ref{free_gyro2}--\ref{free_gyro5}. In the first three cases of this section, the effect of chirality is absent i.e. there are no cross terms connecting $U$ and $V$ in the boundary conditions at $z=1$.

\subsubsection{The case of zero gyricity}
\label{problem1}

When $\Omega=0$, the multi-structure exhibits double eigenfrequencies, as shown in Figures \ref{free_gyro2}--\ref{free_gyro5}. In this case, the body connected to the tip of the beam does not spin and has rotational inertia $I_0$ around the principal transverse directions.

The boundary conditions for $\Omega=0$ are
 \begin{equation*}
  \left(\begin{array}{c}
U^{\prime\prime}(1)\\
V^{\prime\prime }(1)\end{array}\right)=I_0\omega^2\left(\begin{array}{c}U^\prime(1)\\ V^\prime(1)\end{array}\right)\label{da1}
 \end{equation*}
along with (\ref{eqnm2}), whose form is not altered by varying the gyricity.
The conditions above correspond to a body connected at the tip of the beam, which can  translate with respect to, and rotate about,  the principal transverse directions, while not precessing or spinning.


{When $\Omega>0$ the double eigenfrequencies  mentioned above split into pairs. The distance between eigenvalues within these pairs increases as the gyricity is increased, as demonstrated in Figure \ref{free_gyro2}. The same splitting effect was also encountered in \cite{Giorgio1} for a beam with a gyro-hinge.}

\subsubsection{The case of infinite gyricity}
\label{problem2}

When $\Omega \to \infty$, the eigenfrequencies of the structure approach the natural frequencies of a beam with a mass at its tip and a sliding end, which can translate but cannot rotate. Eigenfrequencies linked to this problem are shown in Figure \ref{free_gyro2}(a) as the dashed lines at the right of the figure, and this limit is also achieved in the cases considered in Figures \ref{free_gyro4}--\ref{free_gyro5}.
Indeed, as the spin rate becomes large, the nutation angle of the gyroscopic spinner becomes smaller and this results in small rotation at the  connection between the spinner and  the beam. We note that the lowest branch in Figure \ref{free_gyro2}(a) tends to zero as $\Omega$ is increased, and this limit is only reached when $\Omega \to \infty$.  The limit here corresponds to the trivial mode associated with the sliding end problem and is not an eigenvalue.

The zeros of $X(\omega)$ in (\ref{DR}) and (\ref{XY1}) determine the eigenfrequencies of the problem  (\ref{EBeq}), together with the boundary conditions (\ref{eqnm1}) and
\begin{eqnarray}\label{gam02}
 \left(\begin{array}{c}
U^\prime(1)\\
V^\prime(1)\end{array}\right)={\bf 0\rm}\;, \quad \left(\begin{array}{c}
U^{\prime\prime\prime} (1)\\
V^{\prime\prime\prime} (1)
\end{array}\right)+\alpha \omega^2\left(\begin{array}{c}U(1)\\ V(1)\end{array}\right)={\bf 0\rm}\;.
\end{eqnarray}
These conditions represent a beam with a mass and  a sliding end  at its tip.

\subsubsection{The case of negligibly small moments of inertia}
\label{problem3}

In the limit when all moments of inertia tend to zero, the problem becomes non-chiral ($c=I_0=\gamma=0$), and  the boundary conditions
(\ref{eqnm2}) and (\ref{eqn4n}) reduce to become the second condition in (\ref{gam02}) together with
\begin{eqnarray}
&&\left(\begin{array}{c}U^{\prime\prime}(1)\\ V^{\prime\prime}(1)\end{array}\right)= 
{\bf 0}\;.
\label{ZM}
\end{eqnarray}

We also note that the equation (\ref{DR}) is reduced to the form $Z(\omega)=0$, where the function $Z(\omega)$ is defined in (\ref{Z}).

The corresponding physical  problem represents a beam with a clamped base and a point mass at the upper tip.

\subsubsection{The case of gyro-resonance}
\label{problem4_5}
We say that the elastic chiral system is in the state of a \emph{gyro-resonance} if the gyricity of the spinner $\Omega$  and the radian frequency $\omega$ are related by
\begin{equation}
\label{def}
|\Omega|=\frac{1}{\gamma}\omega\;.
\end{equation}
With this particular choice of the gyricity, the matrix in the right-hand side of (\ref{eqn4n}) becomes degenerate.
In Figure \ref{free_gyro2}, we plot the line corresponding to (\ref{def}).

Two cases of gyro-resonance can be identified by referring to the transcendental equation (\ref{DR}). Using (\ref{def}) and substituting (\ref{XY1a}) into (\ref{DR}) we obtain
\begin{eqnarray}
&& [ \alpha c\beta \omega \sin(\sqrt{\omega}) \sinh(\sqrt{\omega}) - Z(\omega)]\nonumber\\
&&\times [ \alpha c\beta \omega \sin(\sqrt{\omega}) \sinh(\sqrt{\omega}) +2I_0 \omega^{3/2} X(\omega)- Z(\omega)]=0\;.\label{Dcc}
\end{eqnarray}


In the first case of gyro-resonance, the eigenfrequencies  are defined as  zeros of the first factor in (\ref{Dcc}), and the boundary conditions at the upper tip of the beam become (\ref{eqnm2}) and  (\ref{ZM}).

The second case corresponds to zeros of the second factor in (\ref{Dcc}).
 The boundary conditions at $z=1$ take the form of (\ref{eqnm2}) together with
\begin{equation*}
  \left(\begin{array}{c}
U^{\prime\prime}(1)\\
V^{\prime\prime }(1)\end{array}\right)=2I_0\omega^2\left(\begin{array}{c}U^\prime(1)\\ V^\prime(1)\end{array}\right)\;.
\end{equation*}

In both cases,  the moments at the tip of the beam are coupled:
\begin{equation*}\label{couple}
U^{\prime\prime}(1)=\mp \I V^{\prime\prime}(1)
\quad \text{ when } \quad \Omega=\pm\frac{1}{\gamma}\omega\;.
\end{equation*}
The solutions of (\ref{Dcc}) are shown in Figure \ref{free_gyro2}, as intersections of the straight lines (\ref{def}) 
with the curves representing the eigenfrequencies $\omega$ as functions of the gyricity $\Omega$.

\begin{figure}
\includegraphics[width=1\textwidth]{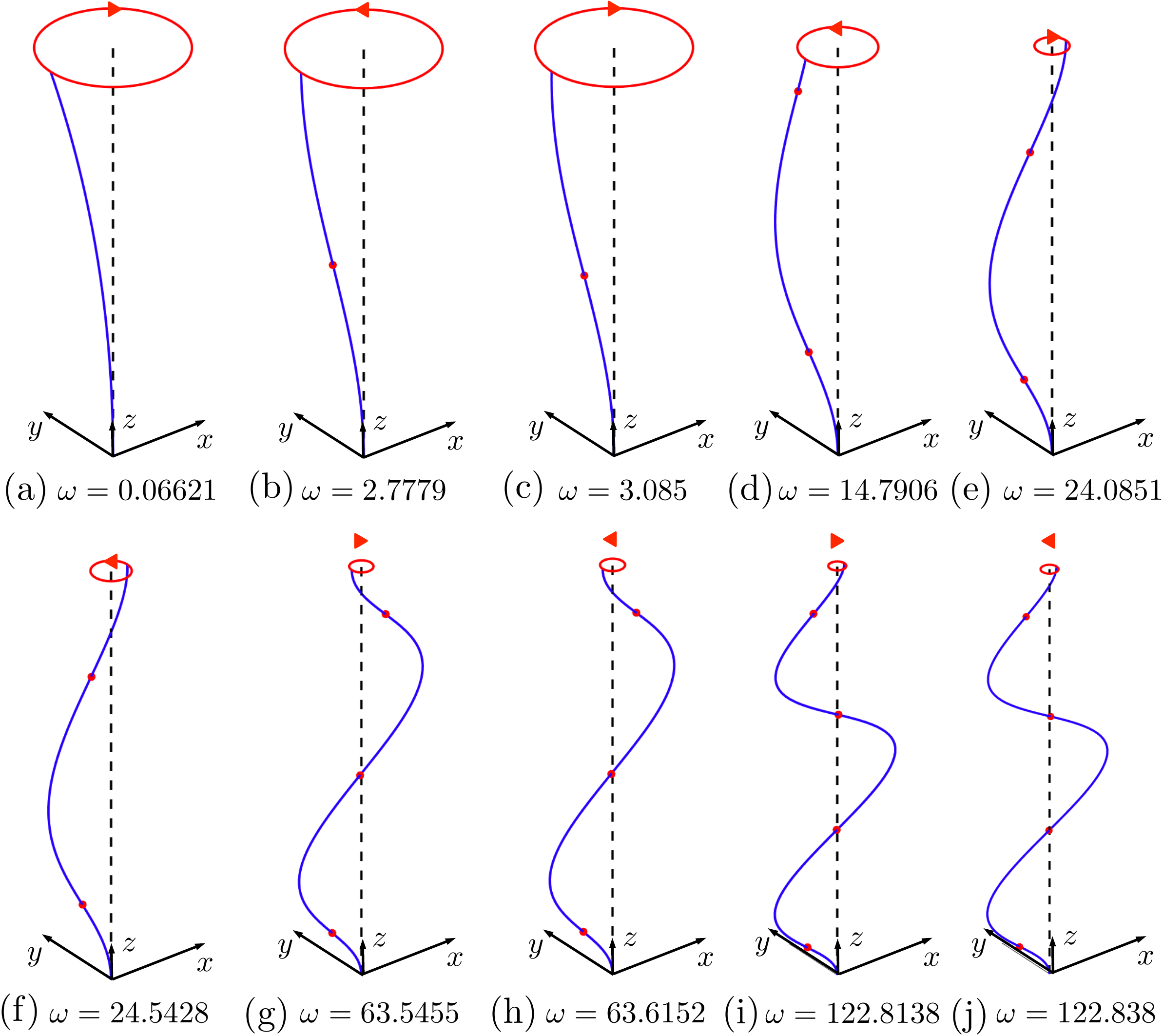}
\caption{Eigenmodes of an inertial beam with a clamped base and a gyroscopic spinner connected to its tip. The computations presented in (a)--(j) are the modes corresponding to the system's eigenfrequencies $\omega$ (indicated in each figure) for a gyricity $\Omega=30$
(see Figure \ref{free_gyro2}(a)).
The beam deformations with respect to the undeformed configuration (dashed lines) are given. The inflection points where the internal moments are zero are shown by dots along the profiles. The trajectory of the tip of the beam during a single period of the system is represented by a circle at the top of each configuration together with an arrow indicating  the direction in which the system rotates. 
}
\label{free_gyro1}
\end{figure}

\subsection{Remarks on chiral waveforms}\label{EManalysis}
We discuss  the eigenmodes corresponding to the  eigenfrequencies obtained for an arbitrarily chosen value of gyricity. In Figure \ref{free_gyro1}, these eigenmodes are computed for the eigenfrequencies corresponding to a gyricity $\Omega=30$
(see Figure \ref{free_gyro2}(a)).

We observe that:
\begin{enumerate}[$\bullet$]
\item with reference to Section \ref{cylindergyro}, 
the first two eigenmodes are asymptotically equivalent to those obtained for the case of the massless  beam connected to the gyroscopic spinner. For these frequency regimes the inertial contribution for the elastic beam is negligibly small.
\item  As the eigenfrequency of the system increases,  the number of inflection points in the beam (where the internal moment is zero) form a monotonic increasing sequence.
\item   The direction of the rotation of the system alternates as the eigenfrequencies of the system increase. For particular eigenfrequencies (stated in Figures \ref{free_gyro1}(f)-(i)),  we observe the time-harmonic motion of the system as the eigenfrequency of the system is increased in Videos {2--5} in the Supplementary Material. The modes in these videos have been  computed assuming $A_1=10^{-5}$ in (\ref{soln}). From these videos it is apparent that the beam rotates in opposite directions in moving from one eigenfrequency of the system to the next largest eigenfrequency, while keeping the gyricity constant the same. In relation to Figure \ref{free_gyro2}(a), {the eigenfrequencies obtained from the monotonic decreasing branches correspond to a clockwise rotation of the system about the undeformed axis of the beam, whereas those associated with the increasing branches provide an anticlockwise rotation.}
\item The displacement of the beam tip from the undeformed configuration decreases as the eigenfrequency of the system increases. It is also apparent from Videos {2-5} in the Supplementary Material that any point along the entire system moves through a circular trajectory as the system completes a period of $2\pi/\omega$, with $\omega$ being the  eigenfrequency of the mode.
\end{enumerate}




\section{Discrete model of a Rayleigh gyrobeam}\label{Periodicst}
The results of the previous sections are used to derive a discrete approximation of a dynamic \emph{Rayleigh gyrobeam}.
The notion of an elastic gyrobeam was introduced  formally in   \cite{DEH1} through a system of differential equations that couple different transverse vibration modes. In the recent paper \cite{Giorgio1}, a periodically constrained gyrobeam was approximated by a discrete elastic system containing so-called gyro-hinges.

In this section,
we consider the discrete approximation of an elastic gyrobeam with an additional rotational inertia. We refer to this  structural element  as a Rayleigh gyrobeam. The approximation is carried out under the assumption
that the variation of the displacements and rotations at the junctions of the system are small.


\subsection{Discrete system of beams connected by gyroscopic spinners}\label{secPS1}
\begin{figure}
\centering
\includegraphics[width=1\textwidth]{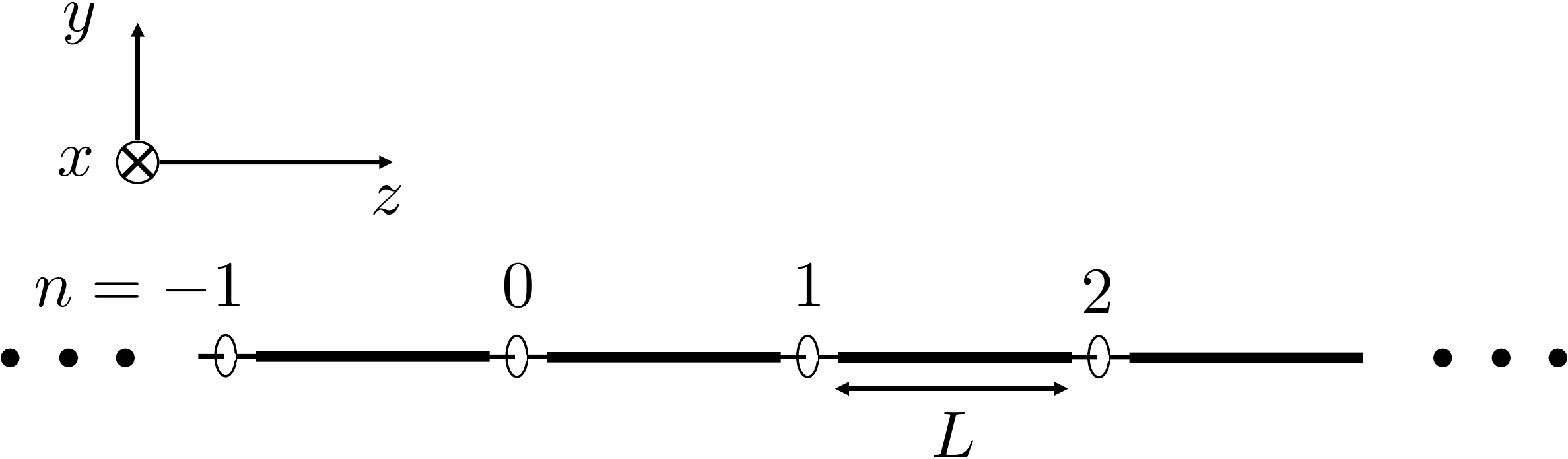}
~~~~~~~~~~~~~~~~~~~~~~~~~(a)

\includegraphics[width=1\textwidth]{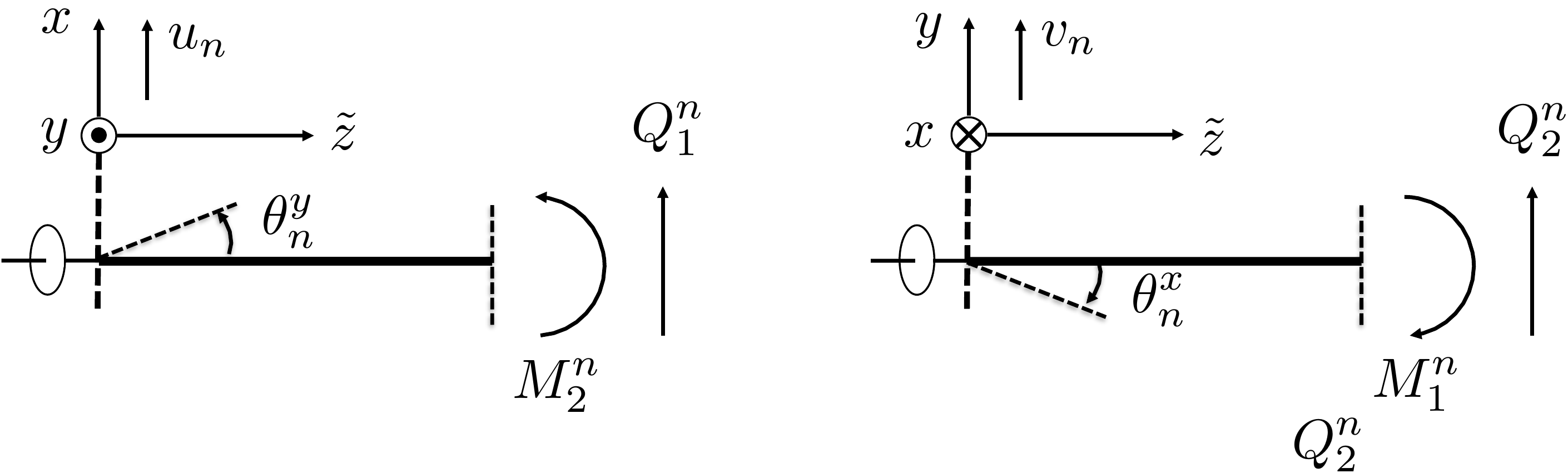}
~~~~~~~~~~~~~~~~~~~~~~~~~(b)
\caption{(a) A structure composed of small gyroscopic spinners, with positions $z=Ln$, $n\in \mathbb{Z}$, connected by massless Euler-Bernoulli beams of length $L$. (b) The positive directions for displacements, rotations, internal bending moments and shear forces in the beam emanating from the \emph{n-{th}} junction. 
}
\label{Fig2}

\end{figure}

We consider the structure shown in Figure \ref{Fig2}(a), which is composed of massless beams connected by small  identical gyroscopic spinners. The spinners are located at $z=nL$, $n\in \mathbb{Z}$, with $L$ being the length of the beams (see Section \ref{chiralbcs}). By a small spinner we mean a spinner possessing mass and moments of inertia, but its length $l$ is small compared to the length of a beam $L$.
In this case,  (i) the  spinners allow for the continuity condition for displacements  between  neighbouring beams  to be employed and (ii)  the moments   generated by the shear forces coming from the beam are small.
For simplicity, we assume that each gyroscopic spinner  possesses the same gyricity $\Omega$ and  the same moment of inertia tensor.
In addition, the gyroscopic spinners do not transmit any spinning motion to neighbouring beams and the inclination of the spinner and rotations of the beams at the junction are the same at any time during the motion.


We introduce a local coordinate $\tilde{z}=z-Ln$ (see Figure \ref{Fig2}(b)).
Here, $U_n$ and $V_n$ are the displacements in the $x$- and  $y$-directions, respectively, in the \emph{n-{th}} beam. 
In the same beam,  let the internal bending moments 
\[ \BM^n=M_1^n\Be_1+M_2^n\Be_2\]
with the components
\begin{eqnarray}
M_1^n=-EJ \frac{\D^2V_n(\tilde{z})}{\D \tilde{z}^2}\;,  \quad M_2^n=EJ \frac{\D^2U_n(\tilde{z})}{\D \tilde{z}^2}\;,\label{MOM_PS}
\end{eqnarray}
and internal shear forces  
\[\BQ^n=Q_1^n\Be_1+Q_2^n\Be_2\] 
with
\begin{eqnarray}
Q_1^n=-EJ \frac{\D^3U_n(\tilde{z})}{\D \tilde{z}^3}\;,  \quad Q_2^n=-EJ \frac{\D^3V_n(\tilde{z})}{\D \tilde{z}^3}\;. \label{SF_PS}
\end{eqnarray}
The functions $U_n$ and $V_n$
in  the \emph{n-th} massless beam  can be written as
\begin{eqnarray}
U_n(z)&=&{[(\theta^y_{n+1}+\theta^y_{n})L -2(u_{n+1}-u_n)]}\frac{{z}^3}{{L^3}}\nonumber\\
&&-[(\theta^y_{n+1}+2\theta^y_{n})L -3(u_{n+1}-u_n)]\frac{{z}^2}{{L^2}}+\theta_n^y {z}+u_n\;,\label{ps1}\\
V_n(z)&=&{-[(\theta^x_{n+1}+\theta^x_{n})L +2(v_{n+1}-v_n)]}\frac{{z}^3}{{L^3}}\nonumber\\
&&+[(\theta^x_{n+1}+2\theta^x_{n})L+3(v_{n+1}-v_n)]\frac{{z}^2}{{L^2}}-\theta_n^x {z}+v_n\;, \label{ps2}
\end{eqnarray}
where at the \emph{n-th} junction of the structure the displacements $u_n$ and $v_n$  are defined as
\[u_n=u(nL, t)=u(nL-0, t)=u(nL+0, t)\;,\]
\[v_n=v(nL, t)=v(nL-0, t)=v(nL+0, t)\;,\]
and  rotations $\theta^x_{n}$ and  $\theta^y_{n}$ are given by
\[ \theta^y_{n}=
\theta_y(nL, t)
=u^\prime (nL, t)=u^\prime (nL-0, t)=u^\prime (nL+0, t)\;, \]
\[ \theta^x_{n}=
\theta_x(nL, t)=
-v^\prime (nL, t)=-v^\prime (nL-0, t)=-v^\prime (nL+0, t)\;.\]
It can be verified with (\ref{ps1}) and (\ref{ps2}) that 
\[U_n(0)= u_n\;, \quad U^\prime_n(0)=\theta_n^y\;, \quad U_n(L)= u_{n+1}\;, \quad U^\prime_n(L)=\theta_{n+1}^y\;, \]
and
\[V_n(0)= v_n\;, \quad V^\prime_n(0)=-\theta_n^x\;, \quad V_n(L)= v_{n+1}\;, \quad V^\prime_n(L)=-\theta_{n+1}^x\;. \]

Using a similar approach to that employed in obtaining   (\ref{J1})--(\ref{J1a}), (\ref{J2a})--(\ref{J2}) and balancing forces and moments about the \emph{n-th} gyroscopic spinner,  we have that the equations governing the linear momentum of the spinner are
\begin{eqnarray*}
&&m \ddot{u}_n=Q_1^n(0)-Q_1^{n-1}(L)\label{ps2aa}\;,\quad m \ddot{v}_n=Q_2^n(0)-Q_2^{n-1}(L)\label{ps2bb}\;,
\end{eqnarray*}
and those determining the rotational motion of the spinner are
\begin{eqnarray*}
&&I_0  \ddot{\theta}^y_n-I_1\Omega \dot{\theta}^x_n
=M_2^n(0)-M_2^{n-1}(L)\;,\\&& I_0  \ddot{\theta}^x_n+I_1 \Omega \dot{\theta}^y_n
=M_1^n(0)-M^{n-1}_1(L)\;.
\label{ps2dd}
\end{eqnarray*}
Then, the above can be rewritten using (\ref{MOM_PS})--(\ref{ps2}), in terms of the quantities  associated with the \emph{n-th} junction in the structure, as
\begin{eqnarray}
m \ddot{u}_n=\frac{6EJ}{L^3}[(\theta^y_{n-1}-\theta_{n+1}^y) L+2(u_{n+1}+u_{n-1}-2 u_n)]\label{ps2a}\;,\\
m \ddot{v}_n=\frac{6EJ}{L^3}[-(\theta^x_{n-1}-\theta_{n+1}^x) L+2(v_{n+1}+v_{n-1}-2 v_n)]\label{ps2b}\;,
\end{eqnarray}
and
\begin{eqnarray}
I_0  \ddot{\theta}^y_n-I_1\Omega\dot{\theta}^x_n
=-\frac{2EJ}{L^2}[(\theta^y_{n+1}+\theta^y_{n-1}+4\theta^y_{n})L-3(u_{m+1}-u_{m-1})]\;, \label{ps2c}\\ \nonumber \\
I_0  \ddot{\theta}^x_n+I_1\Omega \dot{\theta}^y_n =-\frac{2EJ}{L^2}[(\theta^x_{n+1}+\theta^x_{n-1}+4\theta^x_{n})L+3(v_{m+1}-v_{m-1})]\label{ps2d}\;.
\end{eqnarray}

\subsection{Continuum approximation}\label{PS2}

We now assume that the displacements and rotations in the structure vary slowly with respect to the longitudinal  variable $z$ in the structure.
In this case,  we will show below that the behaviour of the effective medium is described by the partial differential equations
\begin{eqnarray}
&&\rho_{\rm e}A\ddot{u}(z, t)+D_{\rm e}u^{{\prime\prime\prime\prime}}(z,t)-J_{\rm e}  \ddot{u}^{\prime\prime}(z, t)-h_{\rm e} \dot{v}^{\prime\prime}(z,t)=0\;,\label{ps3}\\
&&\rho_{\rm e}A\ddot{v}(z, t)+D_{\rm e}v^{{\prime\prime\prime\prime}}(z,t)-J_{\rm e}  \ddot{v}^{\prime\prime}(z, t)+h_{\rm e} \dot{u}^{\prime\prime}(z,t)=0\label{ps4}\;,
\end{eqnarray}
for $z \in \mathbb{R}$, $t>0$, where
\begin{equation*}\label{ps4a1}
\rho_{\rm e}=\frac{m}{LA}\;, \quad D_{\rm e}=EJ\;, \quad J_{\rm e}=\frac{I_0}{L}\quad \text{and} \quad h_{\rm e}=\frac{I_1\Omega}{L}\;
\end{equation*}
represent the effective density, flexural stiffness, {rotational inertia of the beam's  cross-section (about the principal axes)} and gyricity for the medium, respectively. Here $A$ is the cross-sectional area of the beam.

It should be noted that the system (\ref{ps3}), (\ref{ps4}) representing the effective medium for a structure composed of massless beams connected by small and equally spaced gyroscopic spinners, does not represent  an Euler-Bernoulli beam with distributed gyricity as introduced in \cite{DEH1}. In fact, under the assumptions considered here,  this system
describes a \emph{Rayleigh beam} with a constant distribution of gyricity. The Rayleigh beam is a generalisation of Euler-Bernoulli beam theory that incorporates the effect of rotational inertia of the beam's cross section (see the third terms in (\ref{ps3}) and (\ref{ps4})).  These mechanical elements have recently found applications in the design of novel structured flexural materials in \cite{PMRB1, PMRB2, PMRB3}. Further, a system similar to (\ref{ps3})--(\ref{ps4}) has also  been identified in \cite{FQX, ZC} in the analysis of the whirling dynamics of a spinning Rayleigh beam. 

The gyricity constant $h_{\rm e}$ in (\ref{ps3}) and (\ref{ps4}) will be related below to the gyricity $\Omega$ of the spinners. 
When the gyricity $\Omega$ is sufficiently  large, $h_{\rm e}$ becomes sufficiently large  and the chiral terms in (\ref{ps3})--(\ref{ps4}) become dominant compared to the non-chiral terms representing the rotational inertia. For the case of large gyricity the equations (\ref{ps3})--(\ref{ps4}) approximate a classical gyrobeam as in \cite{DEH1}.

 
\vspace{0.1in}
\noindent \vspace{0.1in}
{\bf The homogenised equations}

We derive (\ref{ps3}) and (\ref{ps4}) assuming that the displacements and rotations at the junctions are sufficiently smooth functions  of space and time.

We assume there is a finite interval $L_R={L}/{\varepsilon}$, where $\varepsilon=1/N$ with $N$ being the number of masses in this interval and $0<\varepsilon \ll 1$. In addition, we  assume the displacements and rotations at the junctions vary slowly according to the change in the dimensionless spatial variable  $\hat{z}=\varepsilon n=z/L_R$, so that
 \begin{equation}\label{eqdim1}
 {u}_n={u}_n(\hat{z}, t)\;,\quad {v}_n={v}_n(\hat{z}, t)\;,\quad  \theta^y_n= \theta^y_n(\hat{z}, t)\;, \quad  \theta^x_n= \theta^x_n(\hat{z}, t)\;.
 \end{equation}

Firstly, we  obtain a non-dimensionalised form of (\ref{ps2a})--(\ref{ps2d}).
We introduce the normalisations
\begin{equation}
\label{eqdim2}
u_n=L_R \hat{u}_n, \quad v_n=L_R \hat{v}_n\;, \quad t=\sqrt{\frac{m L_R^3}{EJ}} \hat{t}\;,
\end{equation}
and
\begin{equation}\label{eqdim3}
I_j=mL_R^2 \hat{I}_j\;,\quad  j=0,1\;, \quad \Omega=\sqrt{\frac{EJ}{mL_R^3}}\hat{\Omega}\;.
\end{equation}
Using these normalisations, we obtain from (\ref{ps2a})--(\ref{ps2d}) the following
\begin{eqnarray}
\ddot{u}_n&=&\frac{6}{\varepsilon^3}[(\theta^y_{n-1}-\theta_{n+1}^y) \varepsilon+2(u_{n+1}+u_{n-1}-2 u_n)]\label{Aps2a}\;,\\
 \ddot{v}_n&=&\frac{6}{\varepsilon^3}[-(\theta^x_{n-1}-\theta_{n+1}^x) \varepsilon+2(v_{n+1}+v_{n-1}-2 v_n)]\label{Bps2b}\;,\\
I_0  \ddot{\theta}^y_n-I_1\Omega\dot{\theta}^x_n
&=&-\frac{2}{\varepsilon^2}[(\theta^y_{n+1}+\theta^y_{n-1}+4\theta^y_{n})\varepsilon-3(u_{m+1}-u_{m-1})]\;, \label{Cps2c}\\ \nonumber \\
I_0  \ddot{\theta}^x_n+I_1\Omega \dot{\theta}^y_n &=&-\frac{2}{\varepsilon^2}[(\theta^x_{n+1}+\theta^x_{n-1}+4\theta^x_{n})\varepsilon+3(v_{m+1}-v_{m-1})]\label{Dps2d}\;,
\end{eqnarray}
where the symbol ``$\,\hat{\,}\,$'' has been omitted for ease of notation. The parameter $L_R$ defines the length scale in the structure over which variation of the displacements and rotations occur.
The finite differences appearing in (\ref{Aps2a}) and (\ref{Bps2b}) may be replaced by derivatives of the functions $u$, $v$, $\theta_y$ and $\theta_x$ with respect to the dimensionless spatial variable. This procedure in (\ref{Aps2a}) and (\ref{Bps2b}) yields
\begin{equation}\label{ps5}
\ddot{u}(z, t)=-\frac{12 }{\varepsilon}\{\theta^{\prime}_y(z, t)-u^{\prime\prime}(z, t)\}+\varepsilon \{u^{{\prime\prime\prime\prime}}(z,t)-2\theta^{{\prime\prime\prime}}_y(z, t)\}+O(\varepsilon^3)\;,
\end{equation}
\begin{equation}\label{ps6}
\ddot{v}(z, t)=\frac{12}{\varepsilon}\{\theta^{\prime}_y(z, t)+v^{\prime\prime}(z, t)\}+ 	\varepsilon \{v^{{\prime\prime\prime\prime}}(z,t)+2\theta^{{\rm \prime\prime\prime}}_x(z, t)\}+O(\varepsilon^3)\;,
\end{equation}
whereas in (\ref{Cps2c}) and (\ref{Dps2d})  we obtain
\begin{eqnarray}\label{rotA}
\theta_y(z, t)&=& u^\prime (z, t)-\frac{\varepsilon }{12}[I_0  \ddot{\theta}_y(z, t)-I_1 \Omega \dot{\theta}_x(z,t)]\nonumber\\
&&- \frac{\varepsilon^2}{6}[\theta^{\prime\prime}_y(z,t)-u^{\prime\prime\prime}(z,t)]+O(\varepsilon^4)
\end{eqnarray}
and
\begin{eqnarray}\label{rotB}
\theta_x(z, t)&=&-v^\prime (z, t)-\frac{\varepsilon}{12}[I_0  \ddot{\theta}_x(z,t)+I_1\Omega \dot{\theta}_y(z, t)]\nonumber \\
&&-\frac{\varepsilon^2}{6}[\theta^{\prime\prime}_x(z,t)+v^{\prime\prime\prime}(z,t)]+O(\varepsilon^4)\;,
\end{eqnarray}
where are variables in these relations are dimensionless.
 Substituting the last two relations into (\ref{ps5}) and (\ref{ps6}) gives
\begin{eqnarray}
\ddot{u}(z, t)&=&I_0  \ddot{\theta}^{\prime}_y(z, t)-I_1\Omega \dot{\theta}^{\prime}_x(z,t)\nonumber \\&&+2\varepsilon[\theta^{\prime\prime}_y(z,t)-u^{\prime\prime\prime}(z,t)]-\varepsilon u^{{\prime\prime\prime\prime}}(z,t)+O(\varepsilon^2)\;,\label{eqddu}\\ \nonumber \\
\ddot{v}(z, t)&=&-[I_0  \ddot{\theta}^{\prime}_x(z,t)+I_1\Omega \dot{\theta}^{\prime}_y (z, t)]\nonumber\\&&-{2\varepsilon}[\theta^{\prime\prime}_x(z,t)+v^{\prime\prime\prime}(z,t)]
-\varepsilon v^{{\prime\prime\prime\prime}}(z,t)+O(\varepsilon^2)\;.\label{eqddv}
\end{eqnarray}
Employing (\ref{rotA}) and (\ref{rotB}) once more, we can rewrite the 
equations (\ref{eqddu}) and (\ref{eqddv}) in the form
\[\varepsilon^{-1}\ddot{u}(z, t)+u^{{\prime\prime\prime\prime}}(z,t)-\varepsilon^{-1}I_0 \ddot{\theta}^{\prime}_y(z, t)+\varepsilon^{-1}I_1\Omega \dot{\theta}^{\prime}_x(z,t)+O(\varepsilon)=0\;,\]
\[\varepsilon^{-1}\ddot{v}(z, t)+v^{{\prime\prime\prime\prime}}(z,t)+\varepsilon^{-1} I_0  \ddot{\theta}^{\prime}_x(z, t)+\varepsilon^{-1} I_1\Omega\dot{\theta}^{\prime}_y(z,t)+O(\varepsilon)=0\;.\]
We now take the leading order terms in the above equations  
and substitute  (\ref{rotA}) and (\ref{rotB}). Then, we return to the dimensional variables through (\ref{eqdim1})--(\ref{eqdim3}).  and obtain (\ref{ps3}) and (\ref{ps4}) to leading order. This completes the homogenisation process.

\subsection{Dispersion  of waves in the Rayleigh gyrobeam}\label{secPS3}
\begin{figure}
\centering
\includegraphics[width=0.75\textwidth]{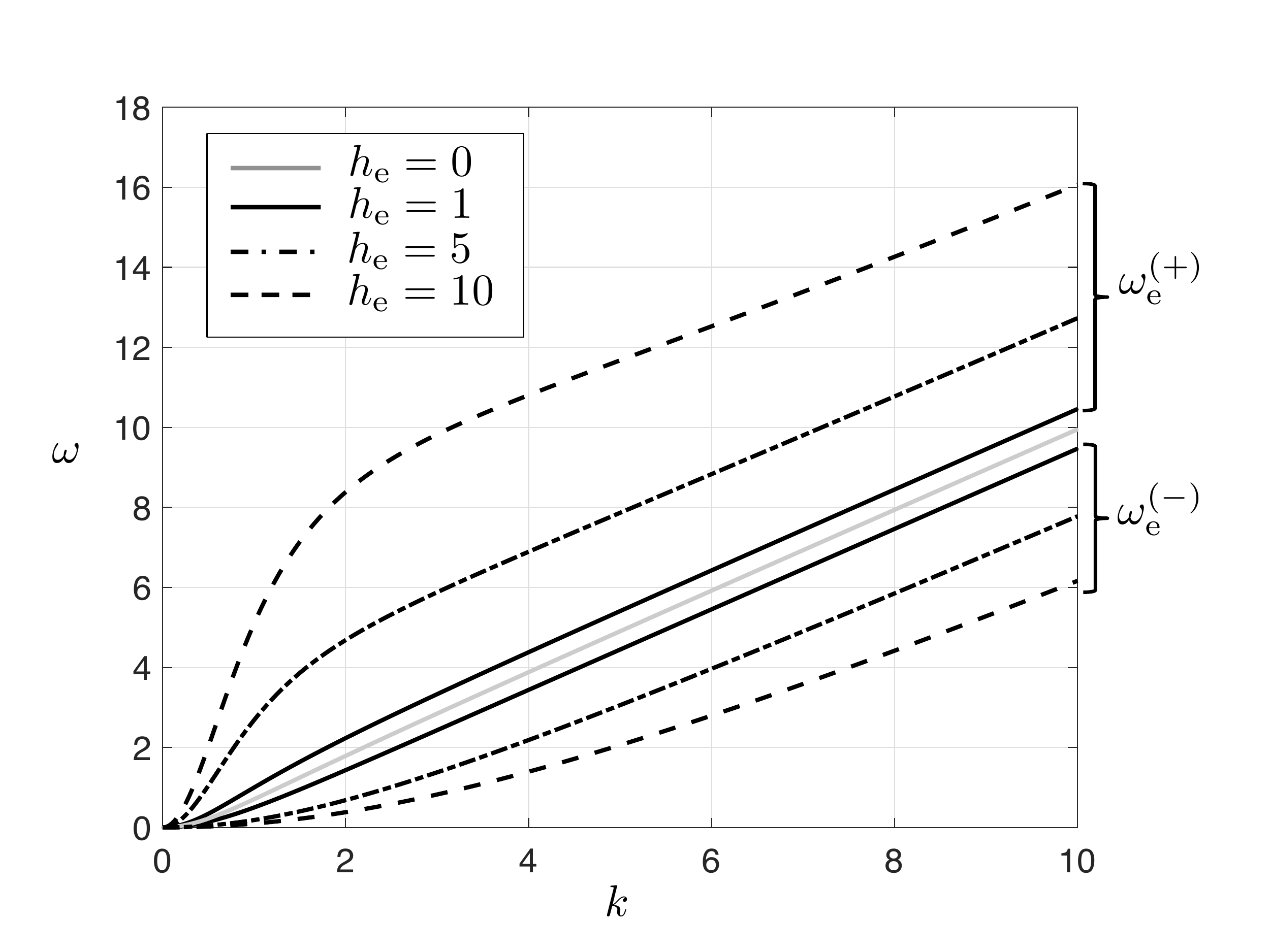}

\caption{Dispersion curves for an infinite Rayleigh gyrobeam, 
for different gyricities $h_{\rm e}$. Computations are based on (\ref{egbd}) for  $\omega_{\rm e}^{(\pm)}$ .
 The solid light grey curve is the dispersion curve for the Rayleigh beam ($h_{\rm e}=0\,{\rm Ns}$). The solid curve is for the case $h_{\rm e}=1\,{\rm Ns}$. The dash-dot curves and dashed curves correspond to  $h_{\rm e}=5\,{\rm Ns}$ and  $h_{\rm e}=10\,{\rm Ns}$, respectively. All other physical parameters are set to unity.
  }
 \label{RGBfig}
\end{figure}

Waves of the form
\begin{equation*}\label{ps7}
u(z, t)={\CC} \E^{\I (\omega t -k z)}\;, \quad v(z, t)={\CD} \E^{\I (\omega t -k z)}\;,
\end{equation*}
propagating through the Rayleigh gyrobeam, can be found by substituting these relations into 
 (\ref{ps3}) and (\ref{ps4}), to obtain the homogeneous system
\begin{equation*}
\left(\begin{array}{ccc}
-\rho_{\rm e} A \omega^2+(D_{\rm e} k^2-J_{\rm e}\omega^2)k^2 & & {\rm i}\omega k^2 h_{\rm e}\\
 -{\rm i}\omega k^2 h_{\rm e} && -\rho_{\rm e} A \omega^2+(D_{\rm e} k^2-J_{\rm e}\omega^2)k^2 
\end{array}\right) \left(\begin{array}{c}\CC\\ \CD\end{array}\right)={\bf 0\rm}\;,
\end{equation*}
for the amplitudes $\CC$ and $\CD$. Non-trivial solutions of this problem then correspond to roots of the determinant of the preceding Hermitian matrix, which are represented by the dispersion relations
\begin{equation}\label{egbd}
\omega_{\rm e}^{(\pm)}=\frac{(\pm h_{\rm e}+\sqrt{4D_{\rm e}(J_{\rm e} k^2+\rho_{\rm e} A)+h_{\rm e}^2 })k^2}{2(J_{\rm e} k^2+\rho_{\rm e} A)}\;.
\end{equation}
In Figure \ref{RGBfig}, we show how the gyricity $h_{\rm e}$ influences the dispersive nature of the Rayleigh beam. It is shown that the presence of gyricity in the Rayleigh beam causes the dispersion curve (representing double eigenfrequencies) for this beam (the light grey solid line) to split into two curves as $k$ is increased from zero. One curve always remains above this line, the other below. In addition, for a given $k$, the distance of these curves from the grey line monotonically increases as  the gyricity increases.
The group velocity of waves in the Rayleigh gyrobeam are bounded by $\sqrt{D_{\rm e}/J_{\rm e}}$ for waves with sufficiently large wavenumber. In fact this bound is uniform for all $k$ for waves associated with the relation $\omega_{\rm e}^{(-)}$.  As follows from (\ref{egbd}), all waves exhibit zero group velocity for $k=0$. 
For small $k$, one can find that the gyricity significantly affects the upper bound for the group velocity of the waves associated with $\omega_{\rm e}^{(+)}$. For example, concentrating on the results produced for $\omega_{\rm e}^{(+)}$, the dashed curve is much steeper than the dash-dot curve  for approximately $0.5\le k \le 1$.  By increasing  the gyricity we can increase the group velocity of waves in the structure.

We note that for small values of $k$ equation (\ref{egbd}) shows that the values of $\omega$ are close to those of the gyrobeam, where the Rayleigh inertia terms are neglected. 

This concludes the comparison between the system of beams and gyroscopic spinners  and a gyrobeam (or a Rayleigh gyrobeam). This is an interesting point raised in the previous work \cite{Giorgio1} where the structure was periodically constrained. The constraints have been removed now and the present illustration shows clearly that   gyrobeams and Rayleigh gyrobeams represent a continuum approximation of multi-structures incorporating linear flexural elements and gyroscopic spinners.
\section{Conclusions}\label{Conclusions}

A new class of chiral boundary conditions has been derived and analysed for elastic multi-structures incorporating elastic beams connected to gyroscopic spinners. A linearised version of this model was obtained  by assuming that the nutation angle of the spinner is small.

An explicit analytical solution has been derived and periodic motions have been identified for transient problems corresponding to massless elastic beams connected to gyroscopic spinners. Furthermore, the case of distributed inertia was studied and the modal analysis of the chiral elastic system was carried out. Several dynamic regimes have been identified including the case of  gyro-resonance, which corresponds to a degeneracy in the chiral boundary conditions.

The study is extended to the case of a periodic elastic structure composed of beams connecting equally spaced gyroscopic spinners. 
In the continuum limit, we obtained a chiral system  approximating a Rayleigh gyrobeam with distributed rotational inertia and gyricity.

We envisage that the present study opens a new pathway for modelling chiral elastic systems incorporating thin elastic solids connected to  gyroscopic spinners. An important part of the study is the derivation and analysis of chiral boundary/junction conditions. Potential applications include the control of elastic waves in multi-scale solids and the design of  earthquake protection systems.


\vspace*{5mm}
\noindent
{\bf Acknowledgements.\rm} M.J.N. gratefully acknowledges the support of the EU H2020 grant MSCA-IF-2016-747334-CAT-FFLAP. G.C., I.S.J., N.V.M. and A.B.M. would like to thank the EPSRC (UK) for its support through Programme Grant no. EP/L024926/1.

\end{document}